\documentclass[notitlepage,english,aps,prb,preprint,tightenlines,floatfix,longbibliography]{revtex4-1}

\usepackage{graphicx}
\usepackage{dcolumn}
\usepackage{bm}
\usepackage{color}
\usepackage[usenames,dvipsnames]{xcolor}
\usepackage[colorlinks=true,urlcolor=black,citecolor=black, linkcolor = black]{hyperref}
\usepackage{amssymb,ulem,amsmath}
\newcommand{\ket}[1]{\left|#1\right\rangle}

\newcommand{\qql}{\textquotedblleft}
\newcommand{\qqr}{\textquotedblright}
\newcommand{\vc}[1]{\bm{\mathrm{#1}}}

\begin{document}

\title{Superfluidity in topologically nontrivial flat bands}

\author{Sebastiano Peotta} 

\affiliation{COMP Centre of Excellence, Department of Applied Physics, Aalto University School of Science, FI-00076 Aalto, Finland}

\author{P\"aivi T\"orm\"a}
\email{paivi.torma@aalto.fi}

\affiliation{COMP Centre of Excellence, Department of Applied Physics, Aalto University School of Science, FI-00076 Aalto, Finland}
\affiliation{Institute for Quantum Electronics, ETH Zurich, 8093 Zurich, Switzerland}

\begin{abstract} 
Topological invariants built from the periodic Bloch functions characterize new phases of matter, such as topological insulators and topological superconductors. The most important topological invariant is the Chern number that explains the quantized conductance of the quantum Hall effect. Here, we provide a general result for the superfluid weight $D_{\rm s}$ of a multiband superconductor that is applicable to topologically nontrivial bands with nonzero Chern number $C$. We find that the integral over the Brillouin zone of the quantum metric, an invariant calculated from the Bloch functions, gives the superfluid weight in a flat band, with the bound $D_{\rm s} \geq |C|$. Thus, even a flat band can carry finite superfluid current, provided the Chern number is nonzero. As an example, we provide $D_{\rm s}$ for the time-reversal invariant attractive Harper-Hubbard model that can be experimentally tested in ultracold gases. In general, our results establish that a topologically nontrivial flat band is a promising  concept for increasing the critical temperature of the superconducting transition.
\end{abstract}

\maketitle

\section{Introduction}
 
An important result of Bardeen-Cooper-Schrieffer (BCS) theory is the relation $T_{\rm c} \propto \exp\left(-\frac{1}{Un_0(E_{\rm F})}\right)\,,$ between the critical temperature of the superconducting transition and the microscopic parameters of a superconductor, such as the coupling constant $U$ of the effective attractive interaction and the density of states at the Fermi energy $n_0(E_{\rm F})$. This result is valid in the limit where the coupling constant $U$ is much smaller than the bandwidth, which is roughly given in a tight-binding approximation by the hopping energy $J$ between neighbouring atomic orbitals. The BCS formula suggests two ways to increase the critical temperature, namely either to enhance the coupling constant $U$ or the density of states $n_0(E_{\rm F})$. Whereas the electron-electron attraction parametrized by $U$ is the result of complicated many-body physics, not yet well-understood in the case of unconventional superconductors, the density of states can be  more easily obtained and engineered in a single-particle framework by means of band structure calculations.

The density of states  at the Fermi energy  $n_0(E_{\rm F})$ is maximal for vanishing bandwidth and so is the critical temperature. In this limit, the energy dispersion as a function of lattice quasi-momentum $\hbar\vc{k}$ is constant  $\varepsilon(\vc{k}) = \bar\varepsilon$ and the corresponding energy band is called a \qql flat band\qqr. The exponential suppression of the critical temperature disappears in the flat-band limit $U/J \gg 1$ since BCS theory predicts~\cite{Khodel:1990,Khodel:1994,Volovik:1991,Heikkila:2011,Kopnin:2011} $T_{\rm c} \propto U n_0(E_{\rm F}) \propto U/J$. This might provide the way to reach the grand goal of room-temperature superconductivity.

A crucial question unaddressed in many works on flat-band superconductors~\cite{Khodel:1990,Khodel:1994,Kopnin:2011,Heikkila:2011,Nandkishore:2012,
Uchoa:2013,Kopnin:2013,Tang:2014,Yudin:2014,Khodel:2015} is whether the superfluid mass density $\rho_{\rm s}$, or better, superfluid weight $D_{\rm s}$ (see below), is nonzero, leading to the Meissner effect and dissipationless transport~\cite{Scalapino:1992,Scalapino:1993} that define superconductivity. Within the single-band effective Hamiltonian approximation~\cite{Grosso_Book,Wannier:1962,Nenciu:1991}, in which only the band dispersion enters, the superfluid weight vanishes ($D_{\rm s} \propto J$) since Cooper pairs localize in the individual lattice sites. Finite superfluid currents can be found in some flat-band systems~\cite{Kopnin:2011b}, but a general theory, connecting the superfluid weight to invariants of the band structure (possibly topological invariants) has not yet been provided. The aim of this work is to answer, at a general level, the crucial question whether superfluidity can exist in a flat band and to explore its possible connections with topological properties of the band.

\begin{figure*}
\includegraphics[scale=0.7]{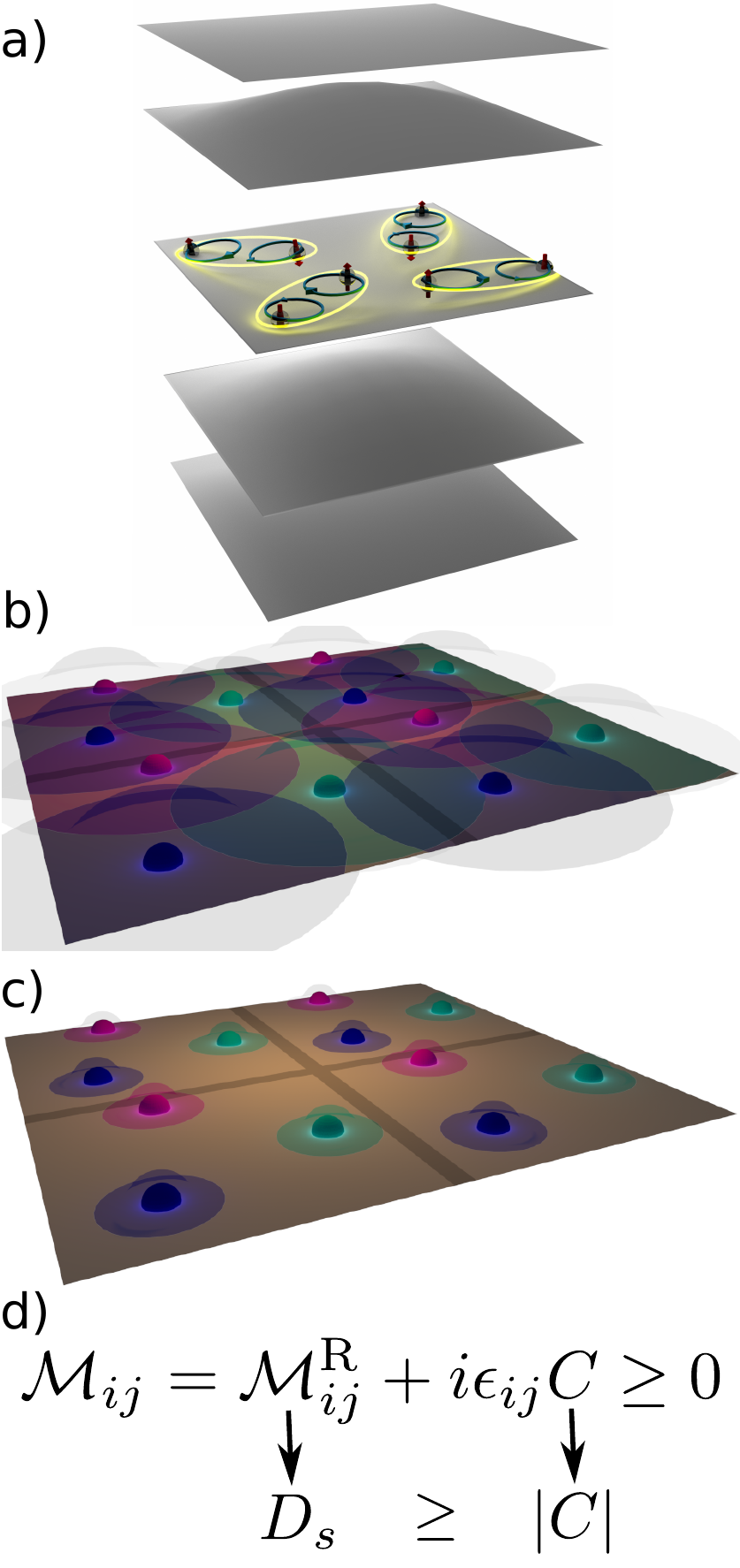}
\caption{\label{fig:one} (\textbf{a}) Localized Wannier functions are obtained from the Bloch functions of a set of bands, called a composite band~\cite{Brouder:2007,Marzari:2012}. To have superfluidity in a flat band, the pairing takes place only in a subset of the bands within the composite band, e.g. in a single flat band.  While the Wannier functions built from the Bloch functions of the band where pairing takes place are delocalized due to the nonzero Chern number\cite{Brouder:2007} $C\neq 0$ (\textbf{b}) the Wannier functions of the composite band are exponentially localized (\textbf{c}). We show that the superfluid weight $D_{\rm s}$ in a flat band is given by the Brillouin-zone average of the quantum metric~\cite{Provost:1980,Berry:1989}, which is the real part of an invariant $\mathcal{M}_{ij}$. The imaginary part of $\mathcal{M}_{ij}$ gives the Chern number $C$. The positivite semidefiniteness of $\mathcal{M}_{ij}$ leads to the bound $D_{\rm s} \geq |C|$ on the superfluid weight.}
\end{figure*}

Using a multiband BCS framework, we show that the superfluid density depends not only on the energy dispersion but also on the Bloch functions of a lattice Hamiltonian. This fact is especially important in the flat-band limit. Moreover, we argue that the superfluid density is subtly affected by the topological invariants encoded in the Bloch functions even in conventional superconductors (not topological)~\cite{Schnyder:2008}. Topological invariants such as the Chern number $C$ are gauge-invariant integer-valued  quantities~\cite{Thouless:1982,Kane:2005} which determine the charge and spin conductance and the presence of robust edge states~\cite{Hasan:2010,Qi:2011,Bernevig_book}.
Indeed, the physical picture of localized Cooper pairs is intimately related to the existence of exponentially localized Wannier functions~\cite{Marzari:2012} that can be constructed only if the Chern number $C$ is nonzero~\cite{Brouder:2007} (see Fig.~\ref{fig:one}).
Note that the Chern number corresponds to an {\it antisymmetric} tensor, the Hall conductance, whereas the superfluid weight is a {\it symmetric} one  and, if nonzero in a flat band, is an invariant quantity constructed only from the Bloch functions. We find that the superfluid density in a flat band is proportional to a symmetric tensor given by the Brillouin-zone average of a quantity known as the quantum metric~\cite{Provost:1980,Berry:1989}. This tensor is the real part of an invariant matrix $\mathcal{M}$, which depends only on the Bloch functions, while the imaginary (antisymmetric) part is the Chern number. By means of the properties of the invariant $\mathcal{M}$, we prove  a bound on the superfluid weight that reads $D_{\rm s} \geq |C|$ in appropriate units (see Fig.~\ref{fig:one}). Moreover, we predict that the superfluid weight is proportional to the coupling constant $D_{\rm s}\propto U$ in a flat band. As a concrete application, we derive the superfluid weight in closed form for the Harper-Hubbard model~\cite{Wang:2014}. Using artificial gauge fields, the Harper model has been recently realized with ultracold gases~\cite{Aidelsburger:2013,Miyake:2013}, which are a good platform to verify our predictions. Our arguments are  general and similar results are expected for other flat bands or bands that are only partially flat.

\section{Results}

\subsection{Effective lattice Hamiltonian}

Our goal is to provide, within a mean-field approximation, a general formula for the superfluid weight of a multiband system which can include topologically nontrivial bands and/or flat bands. A finite supercurrent is associated with a winding of the phase of the superconductor complex order parameter $\Delta(\vc{r})$. In the specific case of a constant current $\vc{J}(\vc{q})$, the order parameter has the form  of a plane wave $\Delta(\vc{r}) = |\Delta|e^{2i\vc{q}\cdot \vc{r}}$ with wavevector $2\vc{q}$. The superfluid mass density $\rho_{\rm s}$ and superfluid weight $D_{\rm s}$ are defined as the change in the free energy density $\Delta F/V = \frac{1}{2}\rho_{\rm s} v_{\rm s}^2 = \frac{1}{8}D_{\rm s} p_{\rm s}^2$ ($V$ is the volume in three dimensions, or the area in two dimensions), due to the motion of Cooper pairs with uniform velocity $v_{\rm s}=\hbar |\vc{q}|/m$ and momentum $p_{\rm s}=2\hbar|\vc{q}|$. In lattice systems, the mass $m$ is not a well-defined concept, and it is better to use the superfluid weight~\cite{Scalapino:1992,Scalapino:1993} $D_{\rm s}$. 

A computationally convenient definition of the superfluid weight is in terms of the grand potential $\Omega(T,\mu,\Delta,\vc{q})$ (see Ref.~\onlinecite{Taylor:2006} and Supplementary Note 1)
\begin{equation}\label{eq:Ds_def}
[D_{\rm s}]_{i,j} = \left.\frac{1}{V\hbar^2}\frac{\partial^2 \Omega}{\partial q_i\partial q_j}\right|_{\mu,\Delta,\vc{q}=0}\,,
\end{equation}
where $i,j = x,y,z$ are spatial indices.
In anisotropic and time-reversal invariant systems, the superfluid weight is given by a symmetric tensor $[D_{\rm s}]_{i,j}$ (the notation $[M]_{i,j}$ for the elements of a matrix $M$, with $i,j$ not necessarily spatial indices, is used throughout the article). 

In calculating the superfluid weight, we proceed in the following way. 1) The supercurrent wavevector $\vc{q}$ is introduced in the Hamiltonian in a way that is rigorous for topologically non-trivial bands: a multiband approach is used. 2) The kinetic Hamiltonian is Fourier transformed, which defines the band dispersions and the Bloch functions. 3) A mean-field approximation is done by introducing a Bogoliubov-de Gennes (BdG) Hamiltonian. 4)  The BdG Hamiltonian is diagonalized to provide a convenient expression for the grand potential. 5) Supercurrent and superfluid weight are obtained as derivatives of the grand potential with respect to $\vc{q}$: the results are given in terms of the band dispersions and the Bloch functions. 6) The results are connected to topological properties of the system.

Some care is needed to introduce the wavevector $\vc{q}$ in the Hamiltonian in a proper way. By a suitable gauge transformation, it is possible to constrain the complex order parameter $\Delta(\vc{r})$ to be real and have the same translational symmetry as the underlying lattice, whereas the wavevector $\vc{q}$ appears in the kinetic term of the lattice Hamiltonian
\begin{equation}\label{eq:Peierls_1}
K_{\vc{i},\vc{j}} \to K_{\vc{i},\vc{j}} e^{i\vc{q}\cdot (\vc{r}_{\vc{i}}-\vc{r}_{\vc{j}})}\, ,
\end{equation}
where the matrix elements $K_{\vc{i},\vc{j}}\propto J$ are hopping amplitudes between lattice sites. If the wavevector $\vc{q}$ is identified with a constant external vector potential  $\vc{A}$ according to $\vc{q} = q\vc{A}/\hbar$, Eq.~(\ref{eq:Peierls_1}) becomes the usual Peierls substitution. 

\begin{figure*}
\includegraphics[scale=0.7]{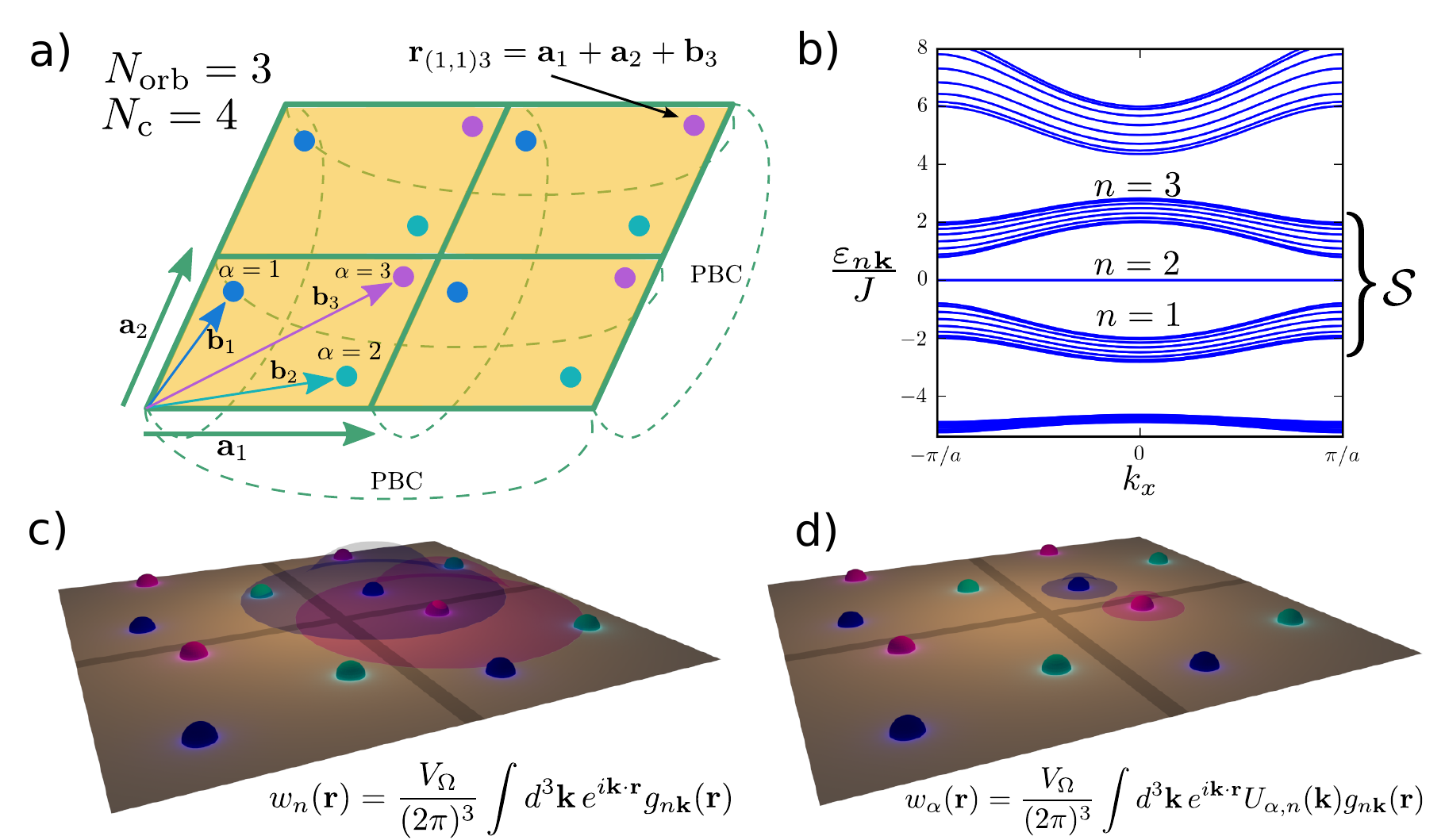}
\caption{\label{fig:two}(\textbf{a}) We consider lattices in three and two dimensions (2D in the figure). Periodic boundary conditions (PBC) are used. The lattice contains $N_{\rm c}$ unit cells and each unit cell contains $N_{\rm orb}$ sites/orbitals. The vectors $\vc{a}_i$ ($i = 1,2,3$) are the fundamental vectors of the Bravais lattice~\cite{Grosso_Book} while the vectors $\vc{b}_\alpha$ ($\alpha = 1,\dots,N_{\rm orb}$) are the positions of the centers of the orbitals (Wannier functions) within a unit cell. A single orbital is specified by a triplet (or pair) of integers $\vc{i} = (i_x,i_y,i_z)$ and by the sublattice index $\alpha$ and is centered at the position vector $\vc{r}_{\vc{i}\alpha} = i_x\vc{a}_1+i_y\vc{a}_2 + i_z\vc{a}_3 + \vc{b}_\alpha$. (\textbf{b}) The band structure is obtained by solving the Schr\"odinger equation $\left(-\frac{\hbar^2\nabla_{\vc{r}}^2}{2m}+V(\vc{r})\right)\psi_{n\vc{k}} = \varepsilon_{n\vc{k}}\psi_{n\vc{k}}$ with periodic potential $V(\vc{r}) = V(\vc{r}+\vc{a}_i)$. It consists of the band dispersions $\varepsilon_{n\vc{k}}$, with $n$ the band index and $\hbar\vc{k}$ the lattice quasimomentum, and the periodic Bloch functions $g_{n\vc{k}}(\vc{r})=g_{n\vc{k}}(\vc{r}+\vc{a}_i)$ (Bloch functions for brevity) obtained from the Bloch plane waves $\psi_{n\vc{k}}(\vc{r}) = e^{i\vc{k}\cdot\vc{r}}g_{n\vc{k}}(\vc{r})$. We consider a composite band, that is, a subset $\mathcal{S}$ of contiguous bands well separated in energy from other bands. The Chern numbers $C_n$ for individual bands calculated from the Bloch functions may be nonzero (such as the flat band $n=2$ in the figure), but their sum equals zero $\sum_{n\in \mathcal{S}}C_n = 0$. The Chern number refers to spin-resolved bands since the spin along a quantization axis (conventionally the $z$ axis) is conserved. (\textbf{c}) The Wannier functions, defined as the Fourier transform of the Bloch functions, allow us to derive a tight-binding Hamiltonian that reproduces exactly a single band or a composite band of the original continuum Hamiltonian (see Supplementary Note 2). Since individual bands may be topologically nontrivial with nonzero Chern numbers, their Wannier functions $w_n(\vc{r})$ are not exponentially localized~\cite{Brouder:2007}, and the Peierls substitution in the effective Hamiltonian is therefore not justified~\cite{Wannier:1962,Nenciu:1991}. (\textbf{d}) By constructing Wannier functions as linear superpositions of Bloch waves of all bands in the composite band, exponentially localized Wannier functions $w_{\alpha}(\vc{r})$  can be created. The mixing of the different bands is provided by the unitary matrix $U_{\alpha,n}(\vc{k})$. This justifies the Peierls substitution for a composite band $\mathcal{S}$.}
\end{figure*}

The Peierls substitution is an approximation valid only if the basis states of the lattice Hamiltonian are well-localized~\cite{Wannier:1962,Nenciu:1991}, and ideally they should be exponentially localized Wannier functions~\cite{Marzari:2012}. Since bands with a nonzero Chern number do not allow exponentially localized Wannier functions~\cite{Brouder:2007}, we use a multiband approach that can circumvent this problem. We consider  a subset of bands, which we call $\mathcal{S}$, well-separated from other bands by band gaps (a composite band)~\cite{Marzari:2012} such that the Chern number (or numbers) of the composite band is zero.
By linear superposition of Bloch functions of all the bands in $\mathcal{S}$, it is possible to construct exponentially localized Wannier functions. For the notation and the definition of Wannier functions, see Figure~\ref{fig:two} and Supplementary Note 2.

In the basis of Wannier functions, the effective lattice Hamiltonian for the composite band reads (the derivation is known~\cite{Grosso_Book}, but for convenience we repeat it in Supplementary Note 2)
\begin{equation}\label{eq:Hamiltonian_start}
\hat{\mathcal{H}} -\mu \hat{N} = \sum_{\vc{i}\alpha,\vc{j}\beta}\sum_\sigma\hat{c}_{\vc{i}\alpha\sigma}^\dagger 
K^\sigma_{\vc{i}\alpha,\vc{j}\beta}e^{i\vc{q}\cdot (\vc{r}_{\vc{i}\alpha}-\vc{r}_{\vc{j}\beta})} \hat{c}_{\vc{j}\beta\sigma} -U\sum_{\vc{i},\alpha,\sigma} \hat{c}_{\vc{i}\alpha\uparrow}^\dagger\hat{c}_{\vc{i}\alpha\uparrow} \hat{c}_{\vc{i}\alpha\downarrow}^\dagger\hat{c}_{\vc{i}\alpha\downarrow} -\mu\hat{N}\,.
\end{equation}
The $\hat{c}_{\vc{i}\alpha\sigma}^{(\dagger)},\hat{c}_{\vc{j}\beta\sigma}^{(\dagger)}$ are annihilation (creation) operators for the orbitals (Wannier functions) labelled by $\vc{i}\alpha$ and $\vc{j}\beta$ (Fig.~\ref{fig:two}\textbf{a}) and spin $\sigma$, $\mu$ the chemical potential, $\hat{N} = \sum_{\vc{i}\alpha\sigma}\hat{c}_{\vc{i}\alpha\sigma}^\dagger\hat{c}_{\vc{i}\alpha\sigma}$ the particle number operator, and we consider the specific case of an attractive Hubbard interaction ($U>0$). The Peierls substitution has been used, properly generalized to the multiband case (see Fig.~\ref{fig:two}\textbf{a} for the definition of $\vc{r}_{\vc{i}\alpha}$). For $\vc{q} = 0$, the Hamiltonian is invariant under time-reversal symmetry (TRS) since $(K^{\uparrow}_{\vc{i}\alpha,\vc{j}\beta})^* = K^\downarrow_{\vc{i}\alpha,\vc{j}\beta}$ and invariant under spin rotation around the $z$-axis, but in general $K^{\uparrow} \neq K^{\downarrow}$.

Diagonalization of the Fourier transform of the hopping matrix in Eq.~(\ref{eq:Hamiltonian_start}) gives the band structure (see Methods~\ref{sec:derivation_BdG})
\begin{gather}\label{eq:band_structure}
\widetilde{K}^\sigma(\vc{k}) = \mathcal{G}_{\vc{k}\sigma}\varepsilon_{\vc{k}\sigma}\mathcal{G}_{\vc{k}\sigma}^\dagger\,.
\end{gather}
Here $\varepsilon_{\vc{k}\sigma} = \mathrm{diag}(\varepsilon_{n\vc{k}\sigma})$ is a diagonal matrix composed of the dispersions $\varepsilon_{n\vc{k}\sigma}$ of each band ($n$ labels a single band belonging to the composite band), while the $n$-th column of the unitary matrix $\mathcal{G}_{\vc{k}\sigma}$ is the  Bloch function $[\mathcal{G}_{\vc{k}\sigma}]_{\alpha,n} = g_{n\vc{k}\sigma}(\alpha)$ of the $n$-th band.  TRS implies that $\varepsilon_{\vc{k}\uparrow} = \varepsilon_{-\vc{k}\downarrow}$ and $\mathcal{G}_{\vc{k}\uparrow}^* = \mathcal{G}_{-\vc{k}\downarrow}$. 

\subsection{BCS theory and superfluid weight in a multiband system}
The idea of superconductivity in multiband systems dates back to~\cite{Suhl:1959} 1959. The first superconductor for which multiband effects are indeed measurable is magnesium diboride (MgB$_2$), discovered as recently as~\cite{Nagamatsu:2001,Xi:2008} 2001. However to the best of our knowledge, a general and consistent theory for the superfluid weight in a multiband system, in particular for topologically nontrivial flat bands, has not yet been worked out.

In the following, we develop the theory of the superfluid weight in a multiband system within the framework of BCS theory, namely, we use a mean-field decoupling of the interaction term
\begin{equation}
-U\sum_{\vc{i},\alpha,\sigma} \hat{c}_{\vc{i}\alpha\uparrow}^\dagger\hat{c}_{\vc{i}\alpha\uparrow} \hat{c}_{\vc{i}\alpha\downarrow}^\dagger\hat{c}_{\vc{i}\alpha\downarrow} \approx \sum_{\vc{i},\alpha} \left(\Delta_{\vc{i}\alpha}\hat{c}_{\vc{i}\alpha\uparrow}^\dagger
\hat{c}_{\vc{i}\alpha\downarrow}^\dagger + \text{H.c.}\right)\,,
\end{equation} 
where $\Delta_{\vc{i}\alpha} = -U\langle \hat{c}_{\vc{i}\alpha\downarrow}\hat{c}_{\vc{i}\alpha\uparrow}\rangle$. 
Furthermore, we choose a gauge where the gap function preserves the discrete translational symmetry of the lattice, which means $\Delta_{\vc{i}\alpha} = \Delta_{\alpha}$. Thus the pairing terms are diagonal in momentum space $\sum_{\vc{i}\alpha}\Delta_\alpha\hat{c}_{\vc{i}\alpha\uparrow}^\dagger
\hat{c}_{\vc{i}\alpha\downarrow}^\dagger= \sum_{\vc{k}\alpha}\Delta_\alpha \hat{c}_{\alpha\vc{k}\uparrow}^\dagger\hat{c}_{\alpha\,-\vc{k}\downarrow}^\dagger$. 
As already discussed, the wavevector $\vc{q}$ enters in the mean-field Hamiltonian in the kinetic energy term through the Peierls substitution~(\ref{eq:Hamiltonian_start}).

In terms of new fermionic operators $\hat{d}_{n\vc{k}\sigma}$ (see Methods~\ref{sec:derivation_BdG}) the mean-field Hamiltonian reads $\hat{\mathcal{H}}_{\rm m.f.} = \sum_{\vc{k}}\hat{\vc{d}}_{\vc{k}}^\dagger H_{\vc{k}}(\vc{q})\hat{\vc{d}}_{\vc{k}}$ with the Nambu spinor given by $\hat{\vc{d}}_{\vc{k}} = (\hat{d}_{n\vc{k}\uparrow},\hat{d}_{n'\,-\vc{k}\downarrow}^\dagger)^T$ and the BdG Hamiltonian is a $2\times 2$ block matrix defined by
\begin{equation}\label{eq:BdG}
H_{\vc{k}}(\vc{q}) = \begin{pmatrix}
\varepsilon_{\vc{k}-\vc{q}}-\mu \bm{1}& \mathcal{G}_{\vc{k}-\vc{q}}^\dagger\Delta\mathcal{G}_{\vc{k}+\vc{q}} \\
\mathcal{G}_{\vc{k}+\vc{q}}^\dagger\Delta\mathcal{G}_{\vc{k}-\vc{q}}  & -\left(\varepsilon_{\vc{k}+\vc{q}}-\mu \bm{1}\right)
\end{pmatrix}\,,
\end{equation}
with $\Delta  = \Delta^* = \mathrm{diag}(\Delta_\alpha)$. 
Due to TRS, the spin index $\sigma$ has been dropped, and only the eigenvalues and eigenstates for the spin up are used in the following, namely $\varepsilon_{\vc{k}} = \varepsilon_{\vc{k}\uparrow}$ and $\mathcal{G}_{\vc{k}} = \mathcal{G}_{\vc{k}\uparrow}$. 
The BdG Hamiltonian is diagonalized in terms of a diagonal matrix of quasiparticle excitation energies $E_{\vc{k}}(\vc{q})$ and a unitary matrix $\mathcal{W}_{\vc{k}}(\vc{q})$
\begin{gather}
H_{\vc{k}}(\vc{q}) = \mathcal{W}_{\vc{k}}(\vc{q})E_{\vc{k}}(\vc{q})\mathcal{W}_{\vc{k}}^\dagger(\vc{q})\,.
\end{gather}
The symmetries of the BdG Hamiltonian for $\vc{q} = 0$ imply that these matrices have the following structure ($E_{n\vc{k}} > 0$, see Supplementary Note 3)
\begin{gather}
E_{\vc{k}}(\vc{q}=0) = 
\begin{pmatrix}
\mathrm{diag}(E_{n\vc{k}}) & 0 \\
0 & -\mathrm{diag}(E_{n\vc{k}})
\end{pmatrix}\,, \label{eq:sol1}\\
\mathcal{W}_{\vc{k}}(\vc{q}=0)=
\begin{pmatrix}
\mathcal{U}_{\vc{k}} & -\mathcal{V}_{\vc{k}} \\
\mathcal{V}_{\vc{k}} & \mathcal{U}_{\vc{k}}
\end{pmatrix}\,.\label{eq:sol2}
\end{gather}
While the kinetic energy terms of the BdG Hamiltonian~(\ref{eq:BdG}) are diagonal in the band index, the pairing terms depend in a complicated way on the Bloch functions and on the order parameters $\Delta_\alpha$ relative to all orbitals. It is interesting to explore the consequences of this nontrivial structure on superfluid transport. A \qql gauge\qqr\ transformation of the Bloch functions given by $\mathcal{G}_{\vc{k}}\to \mathcal{G}_{\vc{k}} \mathcal{\mathcal{A}}_{\vc{k}}$, with $\mathcal{A}_{\vc{k}}$ a unitary matrix subject to the constraint of commuting with the matrix of band dispersions $[\varepsilon_{\vc{k}},\mathcal{A}_{\vc{k}}] = 0$, leaves the BdG Hamiltonian~(\ref{eq:BdG}) unchanged in form while the eigenfunctions~(\ref{eq:sol2}) change accordingly. This freedom in the definition of the Bloch function is the same one preventing a unique definition of Wannier functions~\cite{Marzari:2012}. All observable quantities, such as current and superfluid weight, are necessarily gauge invariant.

At zero temperature, the grand potential is (see Methods~\ref{sec:matrix_not} and Supplementary Note 3)
\begin{equation}\label{eq:grand_pot}
\Omega(\vc{q})  = -\frac{1}{2}\sum_{\vc{k}}\mathrm{Tr}[|H_{\vc{k}}(\vc{q})|] + \dots\,.
\end{equation}
The dots in the above equation represent terms in the grand potential that do not contribute
to the superfluid weight. The superfluid current density is obtained from the first derivative of $\Omega(\vc{q})$
\begin{gather}
\begin{split}
\vc{J}(\vc{q}) &= \frac{-1}{2V\hbar }\sum_{\vc{k}}\mathrm{Tr}\left[\mathrm{sign}(E_{\vc{k}}(\vc{q}))\mathcal{W}_{\vc{k}}^\dagger(\vc{q})
\partial_{\vc{q}}H_{\vc{k}}(\vc{q})
\mathcal{W}_{\vc{k}}(\vc{q})
\right]
\end{split}\nonumber\\
\text{with} \quad -\partial_{\vc{q}}H_{\vc{k}}(\vc{q}) =  \begin{pmatrix}
\partial_{\vc{k}}\varepsilon_{\vc{k}-\vc{q}} & \partial_{\vc{q}}\mathcal{D}_{\vc{k}}(\vc{q}) \\
-\partial_{\vc{q}}\mathcal{D}_{\vc{k}}(-\vc{q}) & \partial_{\vc{k}}\varepsilon_{\vc{k}+\vc{q}}
\end{pmatrix}\,.\label{eq:current}
\end{gather}
The definition $\mathcal{D}_{\vc{k}}(\vc{q}) = - \mathcal{G}_{\vc{k}-\vc{q}}^\dagger\Delta\mathcal{G}_{\vc{k}+\vc{q}}$ has been employed above. Due to the linearity of the trace in Eq.~(\ref{eq:current}), the current splits into two contributions that are separately gauge invariant. We call the first the \qql conventional\qqr\  current, which depends on the group velocity $\partial_{\vc{k}}\varepsilon_{\vc{k}}/\hbar$ and is of order $J/\hbar$, and another contribution of order $\Delta/\hbar$ that comes from the off-diagonal blocks in Eq.~(\ref{eq:current}). Our prediction of the latter current component is highly interesting since it may be nonzero in a flat band, unlike the conventional component.
Note that  in the semiclassical expression for the velocity in a magnetic Bloch band~\cite{Niu:1995,Niu:1996}, two terms appear as well: the group velocity obtained by the band dispersion and the Berry curvature, which is due to interband coupling as the off-diagonal blocks in Eq.~(\ref{eq:current}). However, the analogy is not complete and the precise relation between Berry curvature and the interband contribution to the superfluid density is clarified below.

The superfluid weight is obtained by taking the derivative of the current density $\vc{J}(\vc{q})$ and setting $\vc{q}=0$. The superfluid weight consists of three terms $D_{\rm s} = D_{{\rm s},1}+D_{{\rm s},2}+D_{{\rm s},3}$ (details of the derivation are provided in Supplementary Note 3). We call the first term the conventional superfluid weight
\begin{equation}\label{eq:D1}
[D_{{\rm s},1}]_{i,j} =  \frac{2}{V\hbar^2}\sum_{\vc{k}}\mathrm{Tr}\left[\mathcal{V}_{\vc{k}}\mathcal{V}_{\vc{k}}^\dagger \partial_{k_i}\partial_{k_j}\varepsilon_{\vc{k}}\right]\,.
\end{equation}
This is the only term present in the single band case, and is zero for a flat band. 
The other terms are present only in the multiband case. 
The second term stems from the derivative $\partial_{q_i}$ of the off-diagonal blocks in Eq.~(\ref{eq:current})
\begin{equation}\label{eq:D2}
[D_{{\rm s},2}]_{i,j} = \frac{2}{V\hbar^2}\sum_{\vc{k}}\mathrm{Tr}\left[\mathcal{V}_{\vc{k}}\mathcal{U}_{\vc{k}}^\dagger \partial_{q_i}\partial_{q_j}\mathcal{D}_{\vc{k}}(\vc{q}=0)\right]\,.
\end{equation}
Finally, we have a contribution from terms of the form  $\mathcal{W}_{\vc{k}}^\dagger(\vc{q})\partial_{\vc{q}}\mathcal{W}_{\vc{k}}(\vc{q})$: 
\begin{gather}\label{eq:D3}
[D_{{\rm s},3}]_{i,j} = \frac{2}{V\hbar^2}\sum_{\vc{k}} \sum_{n,n'} \frac{[B_{\vc{k},i}]_{n,n'}[B_{\vc{k},j}]_{n',n}}{E_{n\vc{k}}+E_{n'\vc{k}}}\,, \\
B_{\vc{k},i} = \mathcal{U}_{\vc{k}}^\dagger \partial_{q_i}\mathcal{D}_{\vc{k}}(\vc{q}=0) \mathcal{U}_{\vc{k}} + \mathcal{V}_{\vc{k}}^\dagger \partial_{q_i}\mathcal{D}_{\vc{k}}(\vc{q}=0) \mathcal{V}_{\vc{k}}+ \mathcal{V}_{\vc{k}}^\dagger\partial_{k_i}\varepsilon_{\vc{k}} \mathcal{U}_{\vc{k}} - \mathcal{U}_{\vc{k}}^\dagger\partial_{k_i}\varepsilon_{\vc{k}} \mathcal{V}_{\vc{k}}\,.
\nonumber
\end{gather}
Eqs.~(\ref{eq:D1})-(\ref{eq:D2})-(\ref{eq:D3}) are the main result of our work since the superfluid weight can be readily calculated using only the ground state solution~(\ref{eq:sol1})-(\ref{eq:sol2}). The conventional superfluid weight $D_{{\rm s},1}$ is invariant under gauge transformations, which means that $D_{{\rm s},2}+D_{{\rm s},3}$ is itself gauge invariant, thus the superfluid weight splits into two distinct contributions in the same way as the current.

\subsection{Superfluid weight in a flat band}

The general results in Eq.~(\ref{eq:D1})-(\ref{eq:D3}) can be specialized to the case of a flat band in two dimensions, and a particularly  interesting case is that of a topologically nontrivial flat band. A band specified by $\bar{n}$ within the composite band $\mathcal{S}$ is considered for which the band gaps separating it from the lower ($\bar{n}-1$) and upper ($\bar{n}+1$) bands are large with respect to the bandwidth. It is thus possible to have a coupling constant $U$ such that 
\begin{equation}
\label{eq:weak_coupling_condition}
\begin{split}
&\left.\begin{matrix}\mathrm{min}_{\vc{k}}\varepsilon_{(\bar{n}+1)\vc{k}} 
-\mathrm{max}_{\vc{k}}\varepsilon_{\bar{n}\vc{k}}\\
\mathrm{min}_{\vc{k}}\varepsilon_{\bar{n}\vc{k}} 
-\mathrm{max}_{\vc{k}}\varepsilon_{(\bar{n}-1)\vc{k}}
\end{matrix}\right\} \gg U \gg \mathrm{max}_{\vc{k}}\varepsilon_{\bar{n}\vc{k}}  -\mathrm{min}_{\vc{k}}\varepsilon_{\bar{n}\vc{k}}\,.
\end{split}
\end{equation}
In this limit, the dispersion of the $\bar{n}$-th band can be approximated by its average $\varepsilon_{\bar{n}\vc{k}} \approx \varepsilon_{\bar{n}}$. To proceed, it is assumed that the order parameters $\Delta_\alpha$ relative to each orbital in the unit cell are all equal,  in other words, that $\mathcal{D}_{\vc{k}}(\vc{q}=0) = \Delta\bm{1}$ where $\Delta = \Delta_\alpha$ is now a real scalar.  We can prove this fact rigorously  for the Harper-Hubbard model.

Given this assumption, it is shown in Supplementary Note 4 that an approximate self-consistent solution can be found in the limit (\ref{eq:weak_coupling_condition}) and has the following form. The matrices $[\mathcal{U}_{\vc{k}}]_{n,n'} = u_{n}\delta_{n,n'}$ and $[\mathcal{V}_{\vc{k}}]_{n,n'} = v_{n}\delta_{n,n'}$ are diagonal, while the other relevant quantities are
\begin{gather}
\mu = \varepsilon_{\bar{n}} + Un_{\phi}\left(\nu-\frac{1}{2}\right)\,,\\
\Delta = Un_{\phi} u_{\bar{n}} v_{\bar{n}} = Un_{\phi}\sqrt{\nu(1-\nu)}\,,\\
u_n =
\begin{cases}
0, & 0 \leq n < \bar{n}\\
\sqrt{1-\nu}, & n = \bar{n}\\
1, & n > \bar{n}
\end{cases}
\quad v_n = \sqrt{1-u_n^2}\,,\\
E_{n\vc{k}} = 
\begin{cases}
E_{\bar{n}} = \frac{Un_{\phi}}{2}, & n =\bar{n}\,\\
|\varepsilon_{n\vc{k}} - \mu|, & n \neq\bar{n}\,,
\end{cases}
\end{gather}
with $\nu$ the filling factor of the $\bar{n}$-th band and $n_\phi^{-1} = N_{\rm orb}$ (the number of orbitals; see Fig.~\ref{fig:two}\textbf{a}). This solution depends only on Eq.~(\ref{eq:weak_coupling_condition}) and is in fact generic for any flat band. The only assumption is $\Delta_\alpha=\Delta$.

The above solution can be inserted in the general formulas~(\ref{eq:D1})-(\ref{eq:D2})-(\ref{eq:D3}). The conventional superfluid weight $D_{s,1}$ vanishes in the flat band limit, and the remaining part has the form
\begin{equation}\label{eq:superfluid_flat}
[D_{{\rm s}}]_{i,j} =  [D_{{\rm s},2}]_{i,j} + [D_{{\rm s},3}]_{i,j}= \frac{2Un_\phi}{\pi\hbar^2}\nu(1-\nu)\mathcal{M}_{ij}^{\rm R}\,.
\end{equation}
We thus find in the flat-band limit that the superfluid weight is proportional to $\Delta \propto Un_\phi$. This is consistent with Ref.~\onlinecite{Kopnin:2011b} for the specific case of the flat band of surface states in rhombohedral graphite, however, our theory is much more general and can be applied to a variety of systems. This result has to be contrasted with the one for an ordinary superconductor in a parabolic band $D_{\rm s} = n_{\rm p}/m_{\rm eff} \propto J$ (with $n_{\rm p}$ the total particle density and $m_{\rm eff}$ the effective mass) that can be obtained from Eq.~(\ref{eq:D1}), the only term that survives in the single band case. Therefore, an important prediction is that in a flat-band superconductor, the superfluid weight is linearly dependent on the coupling constant, whereas it is independent from it in  an ordinary superconductor. Interestingly also in superconducting graphene  with the chemical potential tuned at the Dirac point, one has~\cite{Kopnin:2008,Sonin:2010} $D_{\rm s}\propto U$.

The matrix $\mathcal{M}_{ij}^{\rm R} = \mathrm{Re}(\mathcal{M}_{ij})$ is the real part of a Hermitian matrix defined as
\begin{equation}\label{eq:M-def}
\mathcal{M}_{ij} = \frac{1}{2\pi}\int_{\rm B.Z.} d^2{\vc{k}}\,\mathcal{B}_{ij}(\vc{k})
\end{equation}
which is the integral over the whole Brillouin zone of the so-called quantum geometric tensor
\begin{equation}\label{eq:B-def}
\mathcal{B}_{ij}(\vc{k}) = 2\mathrm{Tr}\left[(\partial_{k_i}\bar{\mathcal{G}}_{\vc{k}}^\dagger)( \partial_{k_j}\bar{\mathcal{G}}_{\vc{k}})\right] + 2\mathrm{Tr}\left[\bar{\mathcal{G}}_{\vc{k}}^\dagger (\partial_{k_i}\bar{\mathcal{G}}_{\vc{k}}) \bar{\mathcal{G}}_{\vc{k}}^\dagger (\partial_{k_j}\bar{\mathcal{G}}_{\vc{k}})\right]\,,
\end{equation}
where $\mathcal{\bar G}_{\vc{k}}$ is the projection of $\mathcal{G}_{\vc{k}}$ on the $\bar{n}$-th band (see Methods~\ref{sec:positivity}). In mathematical terms, the quantum geometric tensor is the Fubini-Study metric in the projective manifold of quantum states~\cite{Provost:1980,Berry:1989}. The quantum geometric tensor has been recently related to observable quantities 
in a single-particle context such as the noise current spectrum~\cite{Neupert:2013}, and plays an important role in characterizing bands that can host fractional Chern insulators, namely, lattice generalization of the fractional quantum Hall state~\cite{Dobardzic:2013,Roy:2014}.

 It can be shown that $\mathcal{B}_{ij}(\vc{k})$ is zero if $\mathcal{\bar{G}}_{\vc{k}}$ is a square unitary matrix. The case where $\bar{\mathcal{G}}_{\vc{k}}$ is a square matrix corresponds to superfluid pairing including all the bands of the composite band. 
Consistently, only if a strict subset of the bands in the composite band participate in the pairing, the superfluid weight can be nonzero in the flat-band limit. In contrast in the same limit, the whole composite band, which has zero Chern number,  is a set of  localized orbitals with vanishing hopping. The imaginary part of $\mathcal{B}_{ij}(\vc{k})$ is the well-known Berry curvature, and its integral over the Brillouin zone is the Chern number in two dimensions $\mathrm{Im}(\mathcal{M}_{ij}) = \mathcal{M}^{\rm I}_{ij} = \epsilon_{ij}C$ ($\epsilon_{ij}= - \epsilon_{ji}$ is the Levi-Civita symbol). The Chern number refers to spin-resolved bands since the $z$ component of the spin is a conserved quantity. 

The real part of $\mathcal{B}_{ij}(\vc{k})$ is a Riemannian metric~\cite{Provost:1980,Berry:1989} defined over the Brillouin zone, the so-called quantum metric. In two dimensions, the positive semidefiniteness of the $2\times 2$ matrix $\mathcal{M}_{ij}$ (see Methods~\ref{sec:positivity} and Fig.~\ref{fig:one}) implies that $\mathrm{det}(\mathcal{M}) \geq 0$ or
\begin{equation}\label{eq:bound}
\mathrm{det}(\mathcal{M}^{\rm R}) \geq \mathrm{det}(\mathcal{M}^{\rm I}) = C^2\,.
\end{equation}
For an isotropic system, the matrix $\mathcal{M}^{\rm R}$ is proportional to the identity, and this gives the bound $D_s \geq |C|$, in appropriate units. 
The trace $\mathrm{Tr}\,\mathcal{M}$ is the gauge invariant part of the localization functional for Wannier functions $F$ studied by Marzari and Vanderbilt~\cite{Marzari:1997,Neupert:2013}, pointing to an intimate connection between non-localization of Wannier functions and the superfluid weight. Indeed, Eq.~(\ref{eq:bound}) also implies  that  the localization functional is bounded from below by the Chern number. In 2D, the bound is $F \geq \frac{A_\Omega}{2\pi}|C|$ with $A_\Omega$ the area of the unit cell (see Supplementary Note 5).

\subsection{Time-reversal invariant attractive Harper-Hubbard model}

To make our results more concrete and study the superfluid weight in a quasi-flat band, we consider the specific example of the time-reversal invariant attractive Harper-Hubbard model~\cite{Wang:2014}. This model is defined on a two dimensional square lattice with lattice spacing $a$  by the hopping operator
\begin{equation}\label{eq:Harper}
\begin{split}
K_{\vc{i},\vc{j}}^{\sigma} = -J(&\omega^{-\sigma i_y}\delta_{\vc{i}+\hat{\vc{ x}},{\vc{j}}} + \omega^{\sigma i_y}\delta_{\vc{i}-\hat{\vc{x}},{\vc{j}}} 
+ \delta_{\vc{i}+\hat{\vc{y}},{\vc{j}}} + \delta_{\vc{i}-\hat{\vc{y}},{\vc{j}}})\,,
\end{split}
\end{equation}
with $\hat{\vc{x}} = (1,0)^T$, $\hat{\vc{y}} = (0,1)^T$ and $\omega = e^{2\pi i n_{\phi}}$. The phase factors $\omega^{\pm\sigma i_y}$ are the lattice version of the Landau gauge that introduces a uniform magnetic field with flux per plaquette given by $n_{\phi}$. We consider the case of a commensurate flux $n_{\phi} = 1/Q$ with $Q$ integer.  The magnetic field has opposite signs for opposite spin $\sigma = \uparrow(\downarrow) = \pm$. This guarantees that the Hamiltonian is TRS invariant.  Since $\omega^{Q} = 1$, the discrete translational invariance of the square lattice is broken down to translations by $Q$ lattice sites on the $y$ direction. We can use the previous notation for composite lattices with the relabeling $(i_x,i_y) \to (i_x,\alpha+Qi_y)$. The Bloch functions and band dispersion are solutions of the Harper equation~\cite{Harper:2014} (Supplementary Note 4). 

We are mainly interested in the limit of low flux density per plaquette $n_{\phi} = 1/Q \ll 1$. In this case, the bandwidth of each band is exponentially suppressed with respect to the band gap~\cite{Harper:2014}, thus Eq.~(\ref{eq:weak_coupling_condition}) is satisfied.
As shown in Supplementary Note 4, a self-consistent solution with $\Delta_\alpha = \Delta = \text{const.}$ ($\Delta$ is now a real scalar) can be found, and therefore the result for the superfluid weight in Eq.~(\ref{eq:superfluid_flat}) applies to the Harper-Hubbard model.
The only missing piece is the evaluation of  $\mathcal{M}$ in Eq.~(\ref{eq:M-def})-(\ref{eq:B-def}).
In the low magnetic field limit, a suitable approximation for the Bloch
functions of the lowest bands consistent with the flat-band approximation $\varepsilon_{\bar{n}\vc{k}}\approx \varepsilon_{\bar{n}}$ is\cite{Harper:2014}
\begin{equation}\label{eq:periodic_Bloch}
g_{\bar{n}\vc{k}}(\alpha) \approx \sum_s e^{-ik_y(\alpha-Qs)a} \varphi_{\bar{n}}\left(\alpha-Qs-\frac{Qk_xa}{2\pi}\right)\,,
\end{equation}
where $\varphi_n(\alpha)$ are the eigenfunctions of the harmonic oscillators if $\alpha$ is a continuous variable. In Supplementary Note 4, it is shown that, for the Harper model,
\begin{equation}\label{eq:M-Landau}
\begin{split}
\mathcal{M} &= \begin{pmatrix}
2\bar{n}+1 & -i \\
i & 2\bar{n}+ 1
\end{pmatrix}\,.
\end{split}
\end{equation}
The superfluid weight in the $\bar{n}$-th band (Landau level in the continuum) is proportional to $\propto 2\bar{n}+1$. 
Note how the bound~(\ref{eq:bound})
 is saturated for the lowest Landau level (all Landau levels in the continuum have~\cite{Thouless:1982,Harper:2014} $|C|=1$).
By working directly in the continuum,  we have obtained precisely the result contained in Eq.~(\ref{eq:superfluid_flat}) and~(\ref{eq:M-Landau}) (details are not provided here). More generally, Eqs.~(\ref{eq:superfluid_flat})-(\ref{eq:M-def})-(\ref{eq:B-def}) are valid for a generic flat band, while Eq.~(\ref{eq:M-Landau}) is specific for the Harper Hamiltonian.

\section{Discussion}

We have discovered that an invariant built from the quantum geometric tensor, which is intimately related to the Chern number, governs superfluidity in the flat-band limit. 
The inequality (\ref{eq:bound}) implies that a topologically nontrivial flat band ($C\neq 0$) is guaranteed to have a finite superfluid density in the presence of pairing in the system. Similar but more complicated bounds are also expected in three dimensions, since $\mathcal{M}_{ij}$ is positive semidefinite in general, and its imaginary part encodes three Chern numbers instead of one.
This is the first time that the superfluid weight has been directly related to a topological invariant. Remarkably, BdG Hamiltonians with TRS and invariance under spin rotation around a given axis belong to the chiral unitary class AIII, whose ground state is topologically trivial in 2D according to the classification of Ref.~\onlinecite{Schnyder:2008}, therefore we are referring to bulk superfluid transport and not to transport due to edge modes.

In a flat band, mean-field theory is usually not adequate, however the BCS wavefunction, implicit in the Bogoliubov-de Gennes approach, is the exact ground state in the continuum limit of the Harper-Hubbard model considered here. This can be shown by mapping to the wavefunction of a quantum Hall ferromagnet~\cite{Girvin:1996,Spielman:2000,Joglekar:2001} (see Methods~\ref{sec:exact_landau}). 
Under this mapping, the result given by (\ref{eq:superfluid_flat}) and (\ref{eq:M-Landau}) for the superfluid weight of the Harper-Hubbard model translates into the spin stiffness or, equivalently, the counterflow-current superfluid density of a quantum Hall ferromagnet~\cite{Girvin:1996,Joglekar:2001} with contact repulsive interactions. Whether mean-field theory can describe pairing in flat bands other than Landau levels is an open problem, analogous to the problem of characterizing the bands that can host a fractional Chern insulator~\cite{Dobardzic:2013,Roy:2014}, but considerably less studied. We have checked that dynamical mean-field theory calculations (which treat local fluctuations exactly) for the Harper-Hubbard model are indeed in excellent agreement with mean-field theory in the case of quasi-flat bands~\cite{Vanhala}.

Another problem of mean-field theory in 2D is that the transition to the normal state occurs at the Berezinsky-Kosterlitz-Thouless (BKT) transition temperature $T_{\rm BKT}$, which is related to the superfluid density by a universal relation and is lower than the mean-field critical temperature. 
At half filling, the estimated $T_{\rm BKT}$ is close to the mean-field transition temperature $T_{\rm c}$ (see Supplementary Note 6 and Supplementary Figure 1)
\begin{equation}
T_{\rm c} = \frac{1}{2}\Delta_{T=0} =\frac{Un_\phi}{4} \gtrsim T_{\rm BKT}\,.
\end{equation}
Indeed, we find $T_{\rm BKT} \approx 0.25,\;0.61,\; 0.75\,T_{\rm c}$ for $\bar{n} = 0,1,2$ respectively.

The superfluid weight is a linear response transport coefficient, a ground state property, and it can be calculated exactly if the exact ground state is known~\cite{Joglekar:2001}, as in the case of the Harper-Hubbard model discussed above. As a consequence, it is not is necessary to employ beyond mean-field methods for estimating the superfluid weight~\cite{Evans:1973}. In summary, while the validity of mean-field theory for flat bands is in general an open question, the superfluid weight derived here for the Harper-Hubbard model is exact in the flat-band limit and a good approximation for quasi-flat bands.

In ultracold gases, the atom-atom interaction is tunable, thus these systems are an ideal platform to confirm our prediction that in a flat-band $D_{\rm s}\propto U$.  In fact, it is possible to introduce complex hoppings in a lattice Hamiltonian (Peierls substitution) by Raman dressing~\cite{Jimenez:2012} or lattice shaking~\cite{Struck:2012}. Notably, the Harper model has been recently implemented with ultracold gases~\cite{Aidelsburger:2013,Miyake:2013}. Whereas at the qualitative level superfluidity in ultracold gases is a well-established fact, a quantitative measurement of the superfluid weight has not been easy to perform so far. It has been proposed that the superfluid weight can be measured by an analog of the classic Adronikashvili experiment~\cite{Cooper:2010}, whereas the superfluid fraction of the unitary Fermi gas has been measured by means of second sound~\cite{Sidorenkov:2013}. Moreover, recent transport experiments with ultracold Fermi gases~\cite{Stadler:2012,Krinner:2015} make it realistic to measure quantities like the superfluid weight.  Currently, the main issue in ultracold gas experiments is the excessive heating present in experiments with artificial gauge fields~\cite{Miyake:2013}. Our estimates indicate (see Methods~\ref{sec:temp-estimate}) that superfluidity may be achieved in the future in topologically nontrivial flat bands that can be realized with ultracold atoms. Flat bands have been suggested as a possible mechanism to explain high-$T_{\rm c}$ superconductors~\cite{Khodel:2015,Yudin:2014}, and our results can be used to prove this hypothesis. If our results are generalized to the long-range Coulomb interaction, one more experimental context where they may be important are quantum Hall ferromagnets (c.f.\ the above-mentioned mapping of the superfluid weight [Eqs.(\ref{eq:superfluid_flat}) and (\ref{eq:M-Landau})] to the spin stiffness of a quantum Hall ferromagnet). In fact, a contact interaction is not an acceptable approximation in this case.

Our results can be understood by distinguishing two possible ways to obtain a band of exactly degenerate states. On the one hand, the particles can be confined in states with negligible overlap by high potential barriers, or alternatively localization can occur in overlapping orbits due to (pseudo-)magnetic fields or lattice geometry. In the latter case, the possibility of transport is a non-trivial question. The fact that we find a nonzero superfluid weight in a flat band can be understood by finite overlap of the Cooper pairs, indeed pairing fluctuations  support transport whenever Cooper pairs can be created and destroyed at distinct locations. Somewhat related in a work~\cite{Huber:2010} that focused on condensation rather than superfluidity, an effective Hamiltonian for bosons in a flat band was derived by taking matrix elements of the interaction between overlapping Wannier functions, which produced an effective hopping for the particles. In the work of Provost and Vallee~\cite{Provost:1980}, pointing out for the first time the natural geometric structure present in a manifold of quantum states, it is suggested that macroscopic quantum systems that exhibit collective behaviour might  be those where the quantum metric has direct physical significance, an intuition that has, in some sense, materialized in our results which showed the connection between quantum metric and superfluid weight. It is an intriguing topic for future research to understand whether the pairing fluctuations and macroscopic phase of a superfluid have any connection to the fact that the quantum metric equals the fluctuations in the quantity that generates the path of a quantum state in the manifold~\cite{Provost:1980}.

While the above discussion may help to guide the intuition, the rigorous framework for future work is given by our results on the important role of Wannier functions in superfluid transport. As we have shown, the bound on the superfluid weight translates into a bound on the the localization functional for Wannier functions~\cite{Marzari:1997}. A nonzero Chern number implies that the Wannier functions have algebraically decaying tails~\cite{Marzari:2012}, and this explains the bound $D_{\rm s} \geq |C|$. But the Wannier functions can also be delocalized on a short range only, which is consistent with the fact that the superfluid weight is related to an invariant distinct from the Chern number. In general, we propose (quasi-)flat bands as a viable way to increase the critical temperature in novel superconducting materials, while at the same time preserving the defining properties of superconductors. We expect the invariant $\mathcal{M}$ that controls the superfluid weight in a flat band to play a central role in this research effort.

\section{Methods}

\subsection{Derivation of the Bogoliubov-de Gennes Hamiltonian}
\label{sec:derivation_BdG}

The BdG Hamiltonian in Eq.~(\ref{eq:BdG}) is important for our purposes, and here we clarify its derivation. The hopping matrix has the same discrete translational symmetry as the Bravais lattice, since $K^\sigma_{\vc{i}\alpha,\vc{j}\beta} = K^\sigma_{\alpha,\beta}(\vc{i}-\vc{j})$. By expanding the field operators into plane waves $\hat{c}_{\vc{i}\alpha\sigma} = \frac{1}{\sqrt{N_{\rm c}}}\sum_{\vc{k}}e^{i\vc{k}\cdot\vc{r}_{\vc{i}\alpha}}\hat{c}_{\alpha\vc{k}\sigma}$,  the kinetic term of the Hamiltonian can be block-diagonalized in momentum space
\begin{gather}
\sum_{\vc{i}\alpha,\vc{j}\beta}\hat{c}_{\vc{i}\alpha\sigma}^\dagger 
K^\sigma_{\vc{i}\alpha,\vc{j}\beta} \hat{c}_{\vc{j}\beta\sigma} = 
\sum_{\vc{k},\alpha,\beta} \hat{c}^\dagger_{\alpha\vc{k}\sigma} [\widetilde{K}^\sigma(\vc{k})]_{\alpha,\beta}
\hat{c}_{\beta\vc{k}\sigma} \label{eq:fourier-hopping_first} \\
\label{eq:fourier-hopping}
\text{with} \quad [\widetilde{K}^\sigma(\vc{k})]_{\alpha,\beta}  = \sum_{\vc{i}-\vc{j}}e^{-i\vc{k}\cdot(\vc{r}_{\vc{i}\alpha}-\vc{r}_{\vc{j}\beta})} K^\sigma_{\vc{i}\alpha,\vc{j}\beta}\,.
\end{gather}
It is convenient to introduce a Nambu spinor $\hat{\vc{c}}_{\vc{k}} = (\hat{c}_{\alpha\vc{k}\uparrow},\hat{c}_{\beta\,-\vc{k}\downarrow}^\dagger)$ built out of the operators $\hat{c}_{\alpha\vc{k}\sigma}$ in the plane wave basis (see Eq.~(\ref{eq:fourier-hopping_first})) and write the mean-field Hamiltonian as $\hat{\mathcal{H}}_{\rm m.f.} = \sum_{\vc{k}}\hat{\vc{c}}_{\vc{k}}^\dagger H_{\vc{k}}'(\vc{q})\hat{\vc{c}}_{\vc{k}}$ with
\begin{equation}
H_{\vc{k}}'(\vc{q}) = \begin{pmatrix}
K^\uparrow(\vc{k}-\vc{q})-\mu\bm{1} & \Delta \\
\Delta & -\left(K^\uparrow(\vc{k}+\vc{q})-\mu \bm{1}\right)
\end{pmatrix}\,.
\end{equation}
To cast the mean-field Hamiltonian in Nambu form, we have anticommuted the spin-down creation and annihilation operators in the kinetic energy term and used TRS in the form $\left(K^\downarrow(\vc{k})\right)^* = K^\uparrow(-\vc{k})$. All the c-number terms in the mean-field Hamiltonian have been dropped since they do not affect the superfluid weight (see Supplementary Notes 1 and 3). A further canonical transformation is performed to go from the basis given by the orbitals within a unit cell, labelled by $\alpha,\beta$, to the basis that diagonalizes the kinetic Hamiltonian, that is, the Bloch functions labelled by $n$. More precisely, the transformation reads $\hat{c}_{\alpha\vc{k}\uparrow} = \sum_{n}[\mathcal{G}_{(\vc{k}-\vc{q})\uparrow}]_{\alpha,n}\hat{d}_{n\vc{k}\uparrow}$ and $\hat{c}_{\beta\,-\vc{k}\downarrow}^\dagger = \sum_{n}[\mathcal{G}_{(-\vc{k}-\vc{q})\downarrow}]^*_{\beta,n}\hat{d}_{n\,-\vc{k}\downarrow}^\dagger = \sum_{n}[\mathcal{G}_{(\vc{k}+\vc{q})\uparrow}]_{\beta,n}\hat{d}_{n\,-\vc{k}\downarrow}^\dagger$.
In this way, Eq.~(\ref{eq:BdG}) is obtained.

\subsection{Definition of a generic function of an Hermitian matrix}
\label{sec:matrix_not}

In Eq.~(\ref{eq:grand_pot}) and Eq.~(\ref{eq:current}), the absolute value $|\cdot|$ and the sign function $\mathrm{sign}(\cdot)$ of the BdG Hamiltonian $H_{\vc{k}}(\vc{q})$ are used. In general, a function $f(\cdot)$ of an Hermitian matrix $H=UDU^{\dagger}$, diagonalized by the unitary matrix $U$ and by the real diagonal matrix $D$, is defined as the function of the eigenvalues $f(H) = Uf(D)U^{\dagger}$.

\subsection{Positive semidefiniteness of the quantum geometric tensor}
\label{sec:positivity}

In Eq.~(\ref{eq:B-def}), the projection $\mathcal{\bar G}_{\vc{k}}$  of the unitary matrix $\mathcal{G}_{\vc{k}}$  on the $\bar{n}$-th band is defined by 
\begin{equation}\label{eq:G-bar-def}
[\bar{\mathcal{G}}_{\vc{k}}]_{\alpha,1} = g_{\bar{n}\vc{k}}(\alpha)=[\mathcal{G}_{\vc{k}}]_{\alpha,\bar{n}}\,, \quad 
\bar{\mathcal{G}}^\dagger_{\vc{k}}\bar{\mathcal{G}}_{\vc{k}} = \bm{1}\,, \quad 
\bar{\mathcal{G}}_{\vc{k}}\bar{\mathcal{G}}_{\vc{k}}^\dagger = P_{\bar{n}\vc{k}}\,.
\end{equation}
$P_{\bar{n}\vc{k}}$ is a projection operator, a positive semidefinite ($P_{\bar{n}\vc{k}} \geq 0$) and idempotent ($P_{\bar{n}\vc{k}}^2 = P_{\bar{n}\vc{k}}$) operator. The matrix $\bar{\mathcal{G}}_{\vc{k}}$
is just a column vector in Eq.~(\ref{eq:G-bar-def}), but it can be a rectangular matrix for a group of degenerate flat bands, for example. Since the dispersion is flat, $\bar{\mathcal{G}}_{\vc{k}}$ characterizes the flat band completely. The positivite semidefiniteness of the projector $P_{n\vc{k}}$  and of its complement $\bm{1}-P_{n\vc{k}}$ implies that the matrix $\mathcal{B}_{ij}(\vc{k})$ in Eq.~(\ref{eq:B-def}) is positive semidefinite since it can be written in the form
\begin{equation}
\mathcal{B}_{ij}(\vc{k}) = 2\mathrm{Tr}\left[(\partial_{k_i}\bar{\mathcal{G}}_{\vc{k}}^\dagger)(\bm{1}-P_{\bar{n}\vc{k}})( \partial_{k_j}\bar{\mathcal{G}}_{\vc{k}})\right] \,.
\end{equation}
The invariant matrix in Eq.~(\ref{eq:M-def}) is also positive semidefinite $\mathcal{M}\geq 0$ since it is a linear combination with positive coefficients of the positive semidefinite matrices $\mathcal{B}_{ij}(\vc{k})$. Interestingly, the Berry curvature (${\rm Im}\,\mathcal{B}_{ij}(\vc{k})$) and the 
Chern number of a set of  bands are obtained by adding the respective contributions of all bands in the set, whereas the quantum metric (${\rm Re}\,\mathcal{B}_{ij}(\vc{k})$) is not additive due to the second term in Eq.~(\ref{eq:B-def}), which is real and involves a double sum over the band index.

\subsection{Exactness of the BCS wavefunction}
\label{sec:exact_landau}

The BCS wavefunction can be shown to be the exact ground state of the Harper-Hubbard model in the flat-band limit. To take the flat-band limit of the Harper-Hubbard model, it is necessary to take the limit of low magnetic flux. The problem is mapped into that of particles in the continuum in the presence of a constant magnetic flux (Landau problem). We consider a  general form for the interparticle interaction potential $V(\vc{r}) = (2\pi)^{-2}\int d^2\vc{q} \,v(\vc{q})e^{i\vc{q}\cdot\vc{r}}$ and perform the projection of the interaction term into the $\bar{n}$-th Landau level~\cite{Roy:2014}
\begin{equation}\label{eq:Ham_proj}
\begin{split}
\hat{\mathcal{H}} = \frac{1}{2}\int \frac{d^2\vc{q}}{(2\pi)^2}\,v(\vc{q})L_{\bar{n}}^2\left(\frac{|\vc{q}|^2\ell_B^2}{2}\right)e^{-{|\vc{q}|^2\ell_B^2}{/2}}
\left(\bar{\rho}_{\bar{n}\vc{q}\uparrow}-\bar{\rho}_{\bar{n}\vc{q}\downarrow}\right)^\dagger
\left(\bar{\rho}_{\bar{n}\vc{q}\uparrow}-\bar{\rho}_{\bar{n}\vc{q}\downarrow}\right)\,.
\end{split}
\end{equation}
Here $\ell_B$ is the magnetic length, $L_{n}(x)$ is the $n$-th Laguerre polynomial and $\bar{\rho}_{n\vc{q}\sigma} = \bar{\rho}_{n(-\vc{q})\sigma}^\dagger$ are projected density operators that obey the Girvin-MacDonald-Platzman algebra
\begin{equation}\label{eq:commutator}
[\bar{\rho}_{n\vc{p}\sigma},\bar{\rho}_{n'\vc{q}\sigma'}] = 
2i\sigma\delta_{nn'}\delta_{\sigma\sigma'}
\sin\left(\frac{\vc{p} \wedge\vc{q}\ell_B^2}{2}\right)\bar{\rho}_{n(\vc{p}+\vc{q})\sigma}\,,
\end{equation}
where $\vc{p} \wedge\vc{q} = p_xq_y -p_yq_x$.
In the Landau gauge, the explicit expression for the projected density operators is
\begin{equation}
\bar{\rho}_{n\vc{q}\sigma} = \sum_k \hat{c}_{n(k-q_x)\sigma}^\dagger\hat{c}_{nk\sigma}e^{i\sigma q_yk\ell_B^2}e^{-\frac{i}{2}\sigma q_xq_y\ell_B^2}\,.
\end{equation}
The annihilation (creation) operators $\hat{c}_{nk\sigma}^{(\dagger)}$ are labelled by the Landau level index $n$ and the momentum $k$ along the $x$ direction which is conserved in the Landau gauge. Notice that if $v(\vc{q})\geq 0$, the interaction Hamiltonian~(\ref{eq:Ham_proj}) is repulsive between particles with the parallel spins and attractive between particles with antiparallel spins. It is straightforward to verify that the operator $\bar{\rho}_{\bar{n}\vc{q}\uparrow}-\bar{\rho}_{\bar{n}\vc{q}\downarrow}$ in the Hamiltonian annihilates the BCS wavefunction
\begin{equation}
\ket{\rm BCS} = \prod_k(u+v\hat{c}_{\bar{n}k\uparrow}^\dagger\hat{c}_{\bar{n}(-k)\downarrow}^\dagger)\ket{\varnothing}
\end{equation}
for arbitrary values of $u$ and $v$ and any value of $\vc{q}$, that is, the BCS wavefunction is a zero eigenvector of the Hamiltonian. Normalization requires that $|u|^2+|v|^2=1$. A possible parametrization is $u = \sqrt{\nu}$ and $v = e^{i\phi}\sqrt{1-\nu}$ with $\nu$ the filling and $e^{i\phi}$ and arbitrary phase. Since the Hamiltonian~(\ref{eq:Ham_proj}) is a positive semidefinite operator for $v(\vc{q})\geq 0$, the BCS wavefunction must be the ground state since it is a zero eigenvector. 

An alternative way to interpret this result is well known in the context of quantum Hall physics~\cite{Girvin:1996,Joglekar:2001}. By performing a particle-hole transformation of the form $\hat{c}_{-k\downarrow}\to \hat{c}_{k\downarrow}^\dagger$, the BCS wavefunction is transformed into the wavefunction of a completely polarized ferromagnet
\begin{equation}\label{eq:wave_ferro}
\ket{\rm Ferro} = \prod_k\left( u\hat{c}_{k\downarrow}^\dagger + v \hat{c}_{k\uparrow}^{\dagger} \right)\ket{\varnothing}\,.
\end{equation}
This is a simple Slater determinant where all the states with spin wavefunction $v\ket{\uparrow} + u\ket{\downarrow}$ are occupied. Under the same transformation, the interparticle interaction becomes a repulsive interaction, which is completely isotropic in spin space. It is easy to understand why the wavefunction Eq.~(\ref{eq:wave_ferro}) is the ground state. According to Hund's rule, the interaction energy is minimized if the all the spins are parallel (a consequence of the Pauli exclusion principle), and in a Landau level, there is no kinetic energy cost that prevents a complete alignment. Indeed, this extreme ferromagnetic state has been observed in experiments in the quantum Hall regime~\cite{Girvin:1996}. It is important to note that the $z$ component of the magnetization in the ferromagnetic state is mapped by the particle-hole transformation into the total number of particles on the superconducting side (and vice versa). Therefore, whereas the wavefunction~(\ref{eq:wave_ferro}) is the ground state when a spinful Landau level is half-filled, the BCS wavefunction is the correct ground state for any filling.

In the limit of a contact interaction, the repulsive interaction between particles with parallel spins disappears and one is left with a purely attractive interaction, that is, the continuum limit of the Harper-Hubbard model considered here.

\subsection{Estimate of the critical temperature for the Harper-Hubbard model}
\label{sec:temp-estimate}

To estimate the critical temperature for an actual ultracold gas experiment, we consider fermionic $^6$Li atoms in an optical lattice with a typical wavelength of the laser standing wave $\lambda=1064\,\mathrm{nm} = 2\pi/k$ and the corresponding recoil energy given by $E_{\rm r} = (\hbar k)^2/2m_{^6\mathrm{Li}} = 1.4\,\mu\mathrm{K}$.
The hopping energy scale $J$ can then be estimated
from the approximate formula
\begin{equation}
J = \frac{4}{\sqrt{\pi}}E_{\rm r}\left(\frac{V_0}{E_{\rm r}}\right)^{3/4}\exp\left(-2\sqrt{\frac{V_0}{E_{\rm r}}}\right)\,.
\end{equation}
Using the same ratio $V_0/E_{\rm r}\approx 7$ as in Ref.~\onlinecite{Aidelsburger:2013} between the amplitude $V_0$ of the optical lattice potential and the recoil energy, one obtains $J \approx 70\,\mathrm{nK}$. In Supplementary Note 4 and Supplementary Figure 2, we estimate that in the isolated flat-band approximation for the time-reversal invariant attractive Harper-Hubbard model, the mean-field critical temperature is of the order of $k_{\rm B}T_{\rm c} \approx 0.02 J$, which implies a BKT transition temperature in the order of the nanoKelvin. 
Such a low temperature results just because we wished to be able to use the analytical results derived here, which requires pairing within a single band and thus $U$ needs to be smaller than the gaps to neighboring bands, Eq.(\ref{eq:weak_coupling_condition}). Conceptually the same results can, however, be achieved when several (but not all) flat (or nearly flat) bands of the composite bands participate in pairing, only that the theoretical analysis becomes more involved. Then, the limit on $U$ is relaxed, and $T_{\rm c}$ can be substantially increased.

\bibliographystyle{apsrev4-1}

\vspace{0.5cm}

\noindent\textbf{Acknowledgments}\\
\noindent This work was supported by the Academy of Finland through
its  Centres  of  Excellence  Programme  (2012-2017)  and
under  Project  Nos.  263347,  251748, and  272490,  and
by  the  European  Research  Council  (ERC-2013-AdG-340748-CODE). We acknowledge useful discussions with Jami Kinnunen, Jildou Baarsma and Tuomas Vanhala. We thank Antti Paraoanu for producing the illustrations for figures 1a)-1c) and 2c)-2d).

\vspace{0.5cm}

\noindent\textbf{Authors Contributions}\\
\noindent All authors contributed to all aspects of this work.

\vspace{0.5cm}

\noindent\textbf{Competing financial interests}\\
\noindent The authors declare no competing financial interests.

\newpage

\end{document}


\subsection*{Supplementary Figures}
\begin{figure}[h!]
\begin{tabular}{cc}
\includegraphics[scale=1.0]{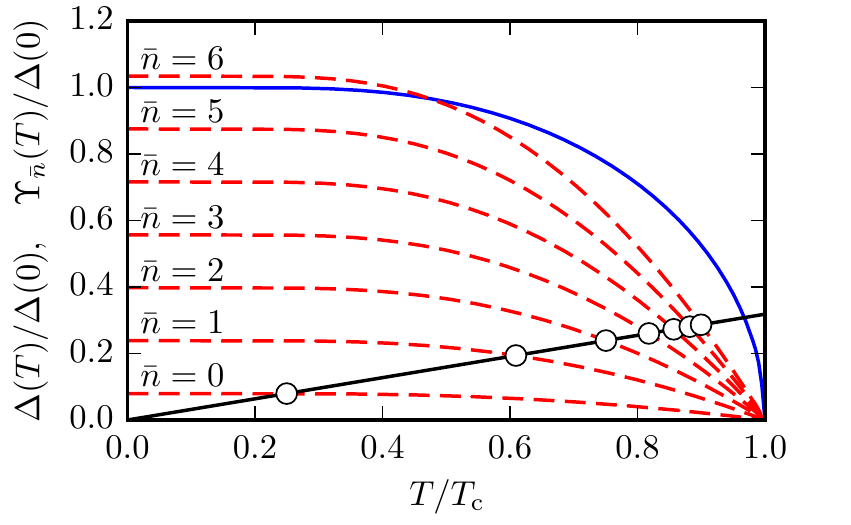} &  \includegraphics[scale=1.0]{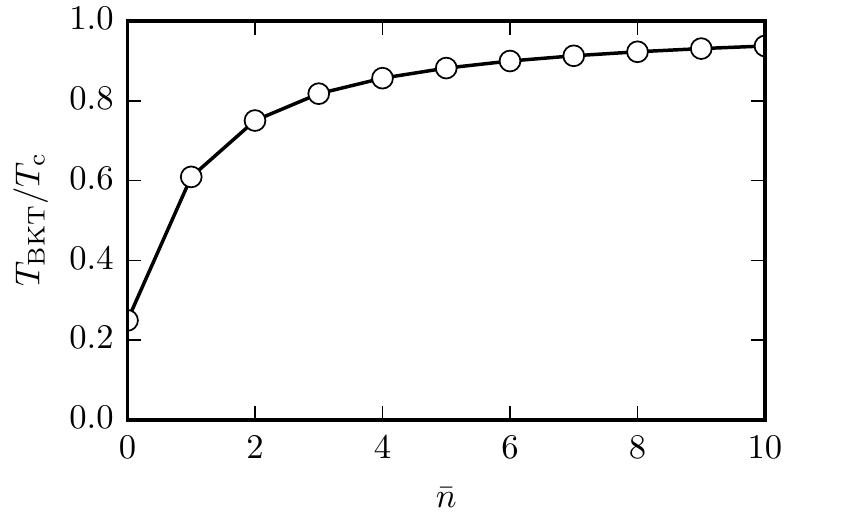}
\end{tabular}
\caption{\label{fig:BKT} On the left plot the energy gap $\Delta(T)/\Delta(0)$ (blue solid line) and the stiffness $\Upsilon_{\bar{n}}(T)/\Delta(0)$ (red dashed line) as a function of temperature are show. The stiffness $\Upsilon_{\bar{n}}$ is shown for several values of $\bar{n}$. The intercepts between the black straight line and the red dashed lines correspond to the solutions of Eq.~(\ref{eq:TBKT}) which gives the BKT temperature $T_{\rm BKT}$ for several values of $\bar{n}$. The calculated solutions of Eq.~(\ref{eq:TBKT}) are shown in the right plot. Only half-filled flat bands ($\nu = 1/2$) are considered here.}
\end{figure}

\begin{figure}[h!]
\includegraphics[scale=1.0]{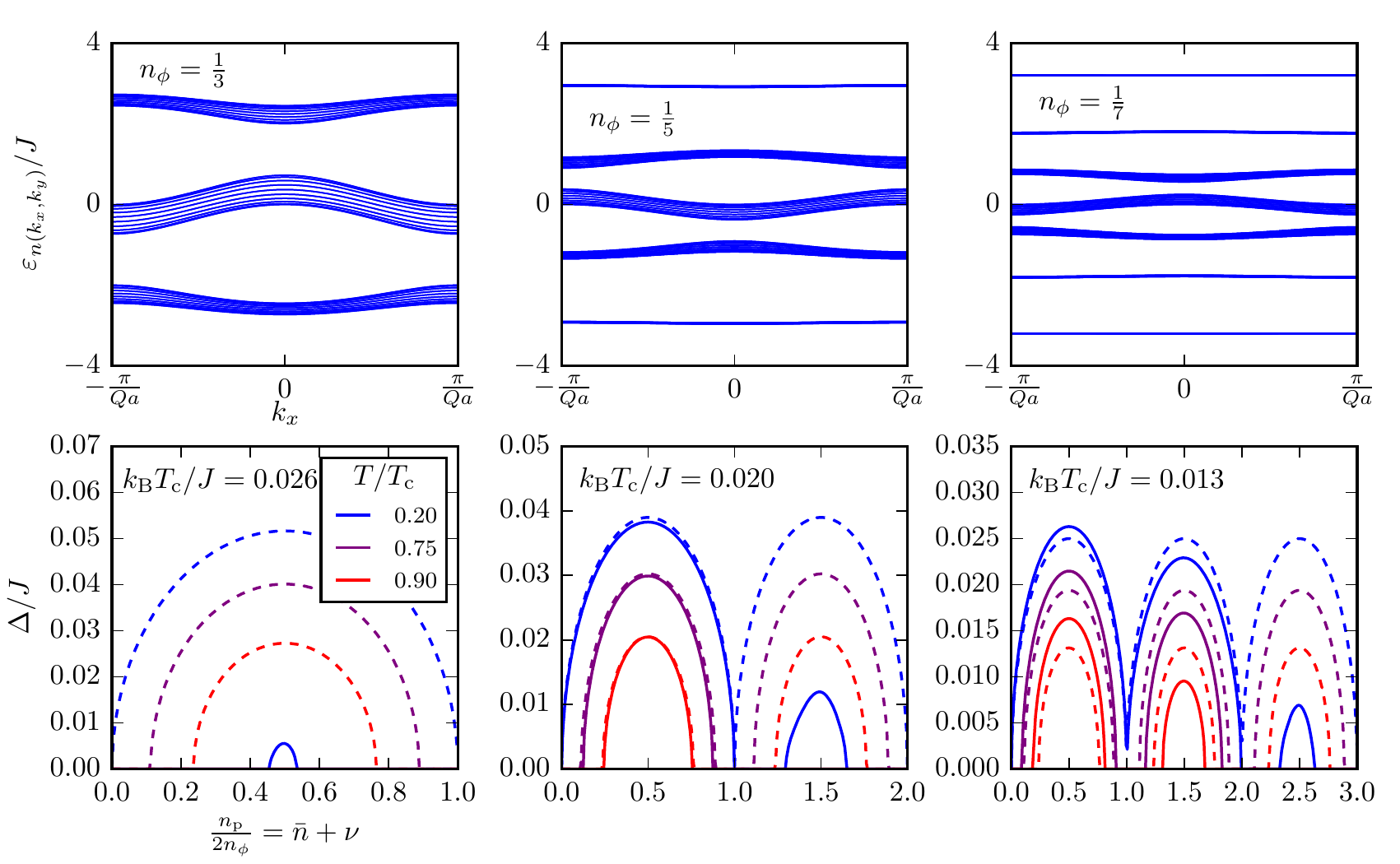}
\caption{\label{fig:Harper_Delta} In the upper panels the band structure of the Harper model for three different values of the number of magnetic flux quanta per plaquette 
$n_\phi = 1/Q = \frac{1}{3},\frac{1}{5},\frac{1}{7}$ is shown. The band dispersion $\varepsilon_{n\vc{k}}$ is shown as a function of the wavevector $k_x$ for twenty different values of $k_y$. Both $k_x,k_y$ are restricted to the reduced Brillouin zone $k_x,k_y \in [-\pi/(Qa),\pi/(Qa)]$ due to the degeneracy of the dispersion in the Harper model~(\ref{eq:Harper-property}). Notice how for decreasing $n_\phi$ the lower bands approach the flat-band limit. 
In the lower panel the  corresponding energy gap $\Delta$ calculated self-consistently by using Eq.~(\ref{eq:gap_eq})-(\ref{eq:mu_eq}) (solid lines) is shown as a function of the normalized total particle density $n_{\rm p}/(2n_\phi) = \bar{n}+\nu$. The value of the coupling constant $U/J$ in all three cases is approximately equal to one-fourth the energy gap between the two lowest bands, more precisely $U/J = 0.31$ for $n_\phi = 1/3$, $U/J = 0.39$ for $n_\phi = 1/5$ and $U/J = 0.35$ for $n_\phi = 1/7$. The critical temperature $k_{\rm B}T_{\rm c} = \frac{Un_\phi}{4} = \Delta_{T=0}/2$ shown in the lower panels is the mean-field critical temperature in the isolated flat-band limit. The numerical self-consistent solution of Eq.~(\ref{eq:gap_eq})-(\ref{eq:mu_eq}) is compared to the approximate solution in the isolated flat-band limit (Eq.~(\ref{eq:Ebar})-(\ref{eq:delta_eq})) (dashed lines) for three different temperatures $T/T_{\rm c} = 0.2,0.75,0.9$. The isolated flat-band approximation is not good for any band in the case $n_\phi = 1/3$, but it can be used for the first band for $n_\phi = 1/5$ and for the first two bands for $n_\phi = 1/7$.} 
\end{figure}

\newpage

\phantom{Aaaaa}

\newpage

\subsection*{Supplementary Note 1: Superfluid weight from the grand potential}\label{sec:mean_field}

The aim of this Supplementary Note is to justify Eq.~(1) in the main text stating that the superfluid weight can be calculated by taking successive derivatives of the grand potential $\Omega(T,\Delta,\mu,\vc{q})$ only with respect to the wavevector $\vc{q}$. The gap function $\Delta(\vc{q})$ and the chemical potential $\mu(\vc{q})$ are themselves a function of $\vc{q}$ (for fixed particle density $n_{\rm p}$). This dependence is in principle important and requires the solution of the self-consistency equations in $\Delta$ and $\mu$ for nonzero values of $\vc{q}$. It is shown here that this not the case. In the following we do not make use of translational invariance since the result that we want to prove holds in general. The greek indices $\alpha,\beta$ that appear in the case of a composite lattice will be suppressed as well since what we are going to prove holds equally well in single-band and multiband systems. The mean-field decoupling of the Hubbard interaction is (including the c-number term)
\begin{equation}\label{eq:mean_field}
\begin{split}
-U\sum_{\vc{j}} {\hat c}_{\vc{j}\uparrow}^\dagger  {\hat c}_{\vc{j}\downarrow}^\dagger {\hat c}_{\vc{j}\downarrow} {\hat c}_{\vc{j}\uparrow}
\approx
\sum_{\vc{j}} \left[ \Delta_{\vc{j}}{\hat c}_{\vc{j}\uparrow}^\dagger  
{\hat c}_{\vc{j}\downarrow}^\dagger + \Delta^*_{\vc{j}}{\hat c}_{\vc{j}\downarrow}  
{\hat c}_{\vc{j}\uparrow} + \frac{|\Delta_{\vc{j}}|^2}{U}\right]\,.
\end{split}
\end{equation}
It will be shown below that the self-consistency equation for the pairing order parameter (also called energy gap function) $\Delta_{\vc{j}} = -U\langle {\hat c}_{\vc{j}\downarrow}  
{\hat c}_{\vc{j}\uparrow}\rangle$ can be obtained as an extremizing condition for the grand potential.

The mean-field Hamiltonian $\mathcal{\hat H}_{\rm m.f.}$, comprising the single particle (quadratic) term in Eq.~(3) in the main text and the mean field approximation~(\ref{eq:mean_field}) for the interaction term  can be conveniently rewritten in matrix form in Nambu space
\begin{equation}\label{eq:nambu}
\begin{split}
{\mathcal{\hat H}}_{\rm m.f.} &= \hat{\vc{c}}^{\dagger} H_{\rm BdG}(\vc{q}) \hat{\vc{c}} + \mathrm{Tr}\,H^\downarrow(\vc{q})+\frac{1}{U}\sum_{\vc{i}}|\Delta_{\vc{i}}(\vc{q})|^2\,,
\end{split}
\end{equation}
where the following definitions have been used
\begin{gather}
\hat{\vc{c}} = \begin{pmatrix}
{\hat c}_{\vc{j},\uparrow} \\ {\hat c}^\dagger_{\vc{j},\downarrow}
\end{pmatrix} \quad (\text{Nambu spinor})\,, \\
H_{\rm BdG}(\vc{q}) = 
\begin{pmatrix}
 H^\uparrow_{\vc{i},\vc{j}}(\vc{q}) & \Delta_{\vc{i}}(\vc{q})\delta_{\vc{i},\vc{j}} \\
\Delta^*_{\vc{i}}(\vc{q})\delta_{\vc{i},\vc{j}} & -H^\downarrow_{\vc{i},\vc{j}}(\vc{q}) 
\end{pmatrix}\,. \label{eq:BdG}\\
H^\uparrow_{\vc{i},\vc{j}}(\vc{q}) = K^\uparrow_{\vc{i},\vc{j}}(\vc{q}) -\mu(\vc{q})\delta_{\vc{i},\vc{j}}+V_{\vc{i}}\delta_{\vc{i},\vc{j}}\,, \label{eq:block_diag1}\\
H^\downarrow_{\vc{i},\vc{j}}(\vc{q}) = \left(K^\downarrow_{\vc{i},\vc{j}}(\vc{q})\right)^* -\mu(\vc{q})\delta_{\vc{i},\vc{j}}+V_{\vc{i}}\delta_{\vc{i},\vc{j}} 
= K^\uparrow_{\vc{i},\vc{j}}(-\vc{q}) -\mu(\vc{q})\delta_{\vc{i},\vc{j}}+V_{\vc{i}}\delta_{\vc{i},\vc{j}}\,,
 \label{eq:block_diag2}
\end{gather}
The trace in Eq.~(\ref{eq:nambu}) comes from the constant term produced when anticommuting fermion operators $\hat{c}_{\vc{i}\sigma}^\dagger \hat{c}_{\vc{j}\sigma}=\delta_{\vc{i},\vc{j}}-\hat{c}_{\vc{j}\sigma}\hat{c}_{\vc{i}\sigma}^\dagger$ ($\delta_{\vc{i},\vc{j}}$ is the Kronecker delta function). In our case we have anticommuted the fermionic operators relative to the down spin.

In the above equations several quantities depend on the wavevector $\vc{q}$. The hopping matrix $K^\sigma_{\vc{i},\vc{j}}(\vc{q}) = K_{\vc{i},\vc{j}} e^{i\vc{q}\cdot (\vc{r}_{\vc{i}}-\vc{r}_{\vc{j}})}$ depends on $\vc{q}$ according to the Peierls substitution, Eq.~(2) in the main text. The pairing potential $\Delta_{\vc{i}}(\vc{q})$ is a function of $\vc{q}$ due to the fact that the mean-field Hamiltonian~(\ref{eq:nambu}) must be solved self-consistently, while the chemical potential depends on $\vc{q}$ since the particle number per lattice site $n_{\rm p}$ is fixed. We have introduced an additional spin-independent scalar potential $V_{\vc{i}}$ in the diagonal terms of $H_{\rm BdG}$ (see Eq.~(\ref{eq:block_diag1})-(\ref{eq:block_diag2})) since our derivation applies to a system without translational invariance. In Eq.~(\ref{eq:block_diag2}) we have used the time-reversal symmetry (TRS) of the hopping term $\big(K^\downarrow_{\vc{i},\vc{j}}(\vc{q})\big)^*=K^\uparrow_{\vc{i},\vc{j}}(-\vc{q})$. Notice that a finite wavevector $\vc{q}$ breaks TRS since it has the effect of inducing a finite supercurrent in the ground state. However TRS connects the Hamiltonians with opposite wavectors $\vc{q}\to -\vc{q}$, in fact under TRS the sign of the supercurrent is reversed.

A basic thermodynamic function we are interested in is the thermodynamic grand potential $\Omega$ defined from the grand partition function $\mathcal{Z}_{\Omega}$
\begin{equation}\label{eq:grand_potential_1}
\begin{split}
\Omega&\big(T,\mu(\vc{q}),\Delta_{\vc{i}}(\vc{q}),\vc{q}\big) = -\frac{1}{\beta}\ln \mathcal{Z}_\Omega
= -\frac{1}{\beta}\ln \mathrm{Tr}[e^{-\beta \mathcal{\hat H}_{\rm m.f.}}] = \mathrm{Tr}\,H_\downarrow(\vc{q})+\sum_{\vc{i}}\frac{|\Delta_{\vc{i}}(\vc{q})|^2}{U} -\frac{1}{\beta}\ln \mathrm{Tr}[e^{-\beta  \hat{\vc{c}}^{\dagger} H_{\rm BdG}(\vc{q}) \hat{\vc{c}}}]\,.
\end{split}
\end{equation}
The grand potential is a function of the wavevector $\vc{q}$ either directly, since $\vc{q}$ appears in the diagonal blocks of $H_{\rm BdG}(\vc{q})$ through the kinetic energy term $K^\sigma(\vc{q})$ [see Eq.~(\ref{eq:BdG})] or indirectly through the quantities $\mu(\vc{q}),\Delta_{\vc{i}}(\vc{q})$. This distinction is important for what follows.
The last term on the right hand side of the above equation can be evaluated by diagonalizing the BdG Hamiltonian. We use the notation
\begin{equation}\label{eq:diagonalization}
H_{\rm BdG}(\vc{q}) = \mathcal{W}(\vc{q})E(\vc{q})\mathcal{W}^\dagger(\vc{q})\,.
\end{equation}
The matrix $\mathcal{W}(\vc{q})$ is unitary, while $E(\vc{q})$ is diagonal and the diagonal elements are the eigenvalues $E_a(\vc{q})$, namely
\begin{equation}\label{eq:eigenvalues}
[E(\vc{q})]_{a,b} = E_{a}(\vc{q})\delta_{a,b} = E_{a}\big(\mu(\vc{q}),\Delta_{\vc{i}}(\vc{q}),
\vc{q}\big)\delta_{a,b}\,.
\end{equation}
The indices $a,b = 1,\dots,2N_{\rm s}$ run over the $2N_{\rm s}$ eigenvalues of the BdG Hamiltonian and $N_{\rm s} = N_{\rm c}N_{\rm orb}$ is the number of lattice sites (see Fig. 2 in the main text).  The square bracket notation $[M]_{a,b}$ denotes the matrix element in row $a$ and column $b$ of the matrix $M$ while $\delta_{a,b}$ is the Kronecker delta.
The eigenvalues $E_a(\vc{q})$ of the BdG Hamiltonian depend on $\vc{q}$ either directly or indirectly in the same way as the grand potential. The notation $E_{a}(\vc{q})$ is just a convenient short-hand for the right hand side of Eq.~(\ref{eq:eigenvalues}).

The canonical transformation of the fermionic operators $\hat{\vc{c}}$
\begin{equation}\label{eq:canonical}
\hat{\vc{c}} = \mathcal{W}(\vc{q})\hat{\vc{\gamma}} \,,
\end{equation}
preserves the anticommutation relations. The operators $\hat{\vc{\gamma}} = (\gamma_1,\dots,\gamma_{2N_{\rm s}})^T$ are therefore annihilation or creation operators of the fermionic quasiparticles that are the elementary excitations above the ground state of the mean-field Hamiltonian. Using the canonical transformation~(\ref{eq:canonical}) the trace on the right hand side of Eq.~(\ref{eq:grand_potential_1}) can be evaluated
\begin{equation}
\mathrm{Tr}[e^{-\beta  \hat{\vc{c}}^{\dagger} H_{\rm BdG}(\vc{q}) \hat{\vc{c}}}] = \mathrm{Tr}[e^{-\beta  \hat{\vc{\gamma}}^{\dagger} E(\vc{q}) \hat{\vc{\gamma}}}] = \prod_a \mathrm{Tr}[e^{-\beta E_a(\vc{q})\hat\gamma_a^\dagger\hat\gamma_a}] =  \prod_a(1+e^{-\beta E_a(\vc{q})})\,.
\end{equation}
The above identity is a consequence of the fact that the number operators $\hat\gamma_a^\dagger \hat\gamma_a$ are mutually commuting and that the trace of a tensor product is the product of traces $\mathrm{Tr}[A\otimes B] = \mathrm{Tr}[A]\mathrm{Tr}[B]$. The grand potential thus reads
\begin{equation}\label{eq:grand_potential_2}
\Omega\big(T,\mu(\vc{q}),\Delta_{\vc{i}}(\vc{q}),\vc{q}\big) = \mathrm{Tr}\,H_\downarrow(\vc{q})+\sum_{\vc{i}}\frac{|\Delta_{\vc{i}}(\vc{q})|^2}{U} -\frac{1}{\beta}\sum_a\ln(1+e^{-\beta E_a(\vc{q})})\,.
\end{equation}

At equilibrium the grand potential attains a minimum, therefore it must be stationary with respect to variations of the pairing potential, namely
\begin{gather}
0 = \frac{\partial \Omega}{\partial \Delta^*_{\vc{i}}} = \frac{\Delta_{\vc{i}}}{U} + \mathrm{Tr}[\hat\rho_{\rm th}\hat{c}_{\vc{i}\downarrow}\hat{c}_{\vc{i}\uparrow}] = \frac{\Delta_{\vc{i}}}{U} + \langle \hat{c}_{\vc{i}\downarrow}\hat{c}_{\vc{i}\uparrow} \rangle\,.\label{eq:self_delta}
\end{gather}
The expectation values $\langle \,\cdot\, \rangle = \mathrm{Tr}[\hat{\rho}_{\rm th}\,\cdot\,]$  
are taken with respect to the equilibrium state described by the thermal density matrix $\hat{\rho}_{\rm th} = e^{-\beta \hat{\mathcal{H}}_{\rm m.f.}}/\mathcal{Z}_{\Omega}$.
Eq.~(\ref{eq:self_delta}) is the equation that guarantees the self-consistency of the mean-field approximation and is written in the most general form valid for an inhomogeneous system. The total particle number $N$ can be fixed using the thermodynamic relation
\begin{equation}\label{eq:self_number}
N = -\frac{\partial \Omega}{\partial \mu} = \sum_{\vc{i}}\langle \hat{c}_{\vc{i}\uparrow}^\dagger\hat{c}_{\vc{i}\uparrow}+
\hat{c}_{\vc{i}\downarrow}^\dagger\hat{c}_{\vc{i}\downarrow}\rangle \,.
\end{equation}
Eqs.~(\ref{eq:self_delta})-(\ref{eq:self_number}) must in general be solved simultaneously by numerical means. Suppose that such a solution has been found for a given $\vc{q}$. 
We now want to show that a self-consistent solution of the time-reversal transformed state with $\vc{q} \to -\vc{q}$ can be immediately obtained according to the transformation
\begin{gather}\label{eq:sym1}
\Delta_{\vc{i}}(\vc{q}) = \Delta_{\vc{i}}^*(-\vc{q}) \,,\\
\mu(\vc{q}) = \mu(-\vc{q})\,.\label{eq:sym3}
\end{gather} 
The proof proceeds by first solving the BdG Hamiltonian $H_{\rm BdG}(-\vc{q})$ assuming the above relations and then showing that indeed the self-consistency equations~(\ref{eq:self_delta}) and~(\ref{eq:self_number}) are satisfied. Eqs.~(\ref{eq:sym1})-(\ref{eq:sym3}) imply that the BdG Hamiltonian has the following symmetry property 
\begin{gather}\label{eq:chiral_sym}
t_{y} H_{{\rm BdG}}(\vc{q}) t_y = -H_{{\rm BdG}}(-\vc{q}) \quad 
\text{with} \quad t_y = \begin{pmatrix}
 0 & -i \\
i & 0 
\end{pmatrix} \otimes \bm{1}_{N_s}
\,.
\end{gather}
The symmetry operator $t_y$ is simply a Pauli matrix that acts on the particle-hole space of the BdG Hamiltonian, namely the matrix blocks in Eq.~(\ref{eq:BdG}), and acts as the identity on the orbital degrees of freedom (lattice sites). For $\vc{q}=0$ this symmetry has the properties
\begin{equation}\label{eq:chiral_sym_3}
t_y H t_y^{-1} = -H \quad t_y t_y^{\dagger} = 1  \quad t_y^2 = 1\,. 
\end{equation}
and is called a \textit{chiral} or \textit{sublattice} symmetry according to the terminology of Ref.~\ncit{Schnyder:2008}{1}, but it is not related to a particular transformation of the lattice in our case. The symmetry (\ref{eq:chiral_sym}) is the result of the particle-hole symmetry which is always present in a BdG Hamiltonian together with time-reversal symmetry (TRS) and invariance with respect to rotations of the spin in the plane perpendicular to the $z$ axis. TRS is broken for $\vc{q} \neq 0$ as discussed before. This symmetry implies that for $\vc{q} = 0$ our system belongs to the AIII (chiral unitary) symmetry class, while for $\vc{q} \neq 0$ the system belongs to the A (unitary) symmetry class\ucit{Schnyder:2008}{1}.

The symmetry in Eq.~(\ref{eq:chiral_sym}) implies that 
\begin{equation}\label{eq:sym_diag}
t_y\mathcal{W}(\vc{q})t_y = \mathcal{W}(-\vc{q})\quad \text{and} \quad t_yE(-\vc{q})t_y = -E(\vc{q})\,.
\end{equation}
It is important to note that the diagonalization of the BdG Hamiltonian $H_{{\rm BdG}}(\vc{q})$ is not uniquely defined and we can choose to permute the eigenvalues according to the transformation $E(\vc{q}) \to P E(\vc{q}) P$ and $\mathcal{W}(\vc{q}) \to \mathcal{W}(\vc{q}) P$ with $P$ a permutation matrix. However if we require the matrix elements of $\mathcal{W}(\vc{q})$ and $E(\vc{q})$ to be continuous and differentiable functions of $\vc{q}$ then the choice~(\ref{eq:sym_diag}) is the only possible one.

In the following we make use of the following additional matrix notations. A diagonal matrix $D = \mathrm{diag}(d_a)$ is completely characterized by the set of its diagonal matrix elements $\{d_a\}_{a = 1,\dots,\mathrm{dim}(D)}$. A generic function $f$ of a diagonal matrix $D = \mathrm{diag}(d_a)$ is taken element by element $f(D) = \mathrm{diag}(f(d_a))$ and a function of a Hermitian matrix $H = U D U^\dagger $, with $U$ the unitary matrix that diagonalizes $H$, is defined by $f(H) = Uf(D)U^\dagger$. In particular we use the modulus of an Hermitian matrix $|H| = U|D|U^\dagger$ and the sign function $\mathrm{sign}(H) = U\,\mathrm{sign}(D)\,U^\dagger$. The sign function is defined in a such a way that $\mathrm{sign}(H)|H| = H$.

We can use the above identities~(\ref{eq:sym_diag}) to calculate the expectation values of bilinear combinations of the fermionic operators $\hat{c}_{\vc{i}\sigma},\hat{c}_{\vc{i}\sigma}^\dagger$ which can be conveniently organized into a matrix
\begin{equation}\label{eq:expectations}
\begin{split}
\langle \hat{\vc{c}} \otimes \hat{\vc{c}}^\dagger \rangle_{-\vc{q}} &= 
\begin{pmatrix}
\langle \hat{c}_{\vc{i}\uparrow}\hat{c}_{\vc{j}\uparrow}^\dagger \rangle_{-\vc{q}}&  \langle \hat{c}_{\vc{i}\uparrow}\hat{c}_{\vc{j}\downarrow} \rangle_{-\vc{q}} \\[1em]
\langle \hat{c}_{\vc{i}\downarrow}^\dagger\hat{c}_{\vc{j}\uparrow}^\dagger \rangle_{-\vc{q}} &
\langle \hat{c}_{\vc{i}\downarrow}^\dagger\hat{c}_{\vc{j}\downarrow} \rangle_{-\vc{q}} 
\end{pmatrix} = 
\mathcal{W}(-\vc{q}) \langle \hat{\vc{\gamma}}\otimes \hat{\vc{\gamma}}^\dagger\rangle_{\vc{-\vc{q}}}\mathcal{W}^\dagger(-\vc{q}) = \mathcal{W}(-\vc{q}) 
\frac{1}{e^{-\beta E(-\vc{q})}+1}\mathcal{W}^\dagger(-\vc{q}) \\ &= \bm{1} - t_y\mathcal{W}(\vc{q}) 
\frac{1}{e^{-\beta E(\vc{q})}+1}\mathcal{W}^\dagger(\vc{q})t_y = \bm{1} -t_y \langle \hat{\vc{c}} \otimes \hat{\vc{c}}^\dagger \rangle_{\vc{q}} t_y = \begin{pmatrix}
\langle \hat{c}_{\vc{j}\downarrow}\hat{c}_{\vc{i}\downarrow}^\dagger \rangle_{\vc{q}}&  \langle \hat{c}_{\vc{i}\downarrow}^\dagger\hat{c}_{\vc{j}\uparrow}^\dagger \rangle_{\vc{q}} \\[1em]
\langle \hat{c}_{\vc{i}\uparrow}\hat{c}_{\vc{j}\downarrow} \rangle_{\vc{q}} &
\langle \hat{c}_{\vc{j}\uparrow}^\dagger\hat{c}_{\vc{i}\uparrow} \rangle_{\vc{q}} 
\end{pmatrix}\,.
\end{split}
\end{equation}
Since the quasiparticle operators $\hat{\gamma}_a$ diagonalize the BdG Hamiltonian $H_{\rm BdG}(\vc{q})$ the matrix $\langle \hat{\vc{\gamma}}\otimes \hat{\vc{\gamma}}^\dagger\rangle_{\vc{q}}$ is diagonal
\begin{equation}
\langle \hat{\vc{\gamma}}\otimes \hat{\vc{\gamma}}^\dagger\rangle_{\vc{q}} = 
\mathrm{diag}\left(\frac{1}{e^{-\beta E_a(\vc{q})}+1}\right) = \frac{1}{e^{-\beta E(\vc{q})}+1}\,.
\end{equation}
Specializing the result in Eq.~(\ref{eq:expectations}) for  $\vc{i} = \vc{j}$  one obtains
Eqs.~(\ref{eq:sym1}).  Moreover since the total particle number $N = \sum_{\vc{i}}(n_{\vc{i}\uparrow}+n_{\vc{i}\downarrow})=\sum_{\vc{i}}\langle \hat{c}_{\vc{i}\uparrow}^\dagger\hat{c}_{\vc{i}\uparrow}+
\hat{c}_{\vc{i}\downarrow}^\dagger\hat{c}_{\vc{i}\downarrow}\rangle $ is unchanged also the chemical potential $\mu(\vc{q})$ is an even function, validating Eq.~(\ref{eq:sym3}). This means that from a self-consistent solution for a certain value of $\vc{q}$  we can immediately provide a self-consistent solution for the TRS conjugate ground state $\vc{q} \to -\vc{q}$. Eq.~(\ref{eq:expectations}) shows that in principle it is possible to have a nonzero spin density in a state with finite current. From Eq.~(\ref{eq:expectations}) the density of the $z$ component of the spin $S_z(\vc{q}) = n_{\vc{i}\uparrow}(\vc{q})-n_{\vc{i}\downarrow}(\vc{q}) = \langle \hat{c}_{\vc{i}\uparrow}^\dagger\hat{c}_{\vc{i}\uparrow}\rangle_{\vc{q}}-\langle
\hat{c}_{\vc{i}\downarrow}^\dagger\hat{c}_{\vc{i}\downarrow}\rangle_{\vc{q}}$ is an odd function of $\vc{q}$, which is consistent with TRS. 

The properties~(\ref{eq:sym1})-(\ref{eq:sym3}), imply that the grand potential is an even function of $\vc{q}$. In fact, consider only the first and the last term in Eq.~(\ref{eq:grand_potential_2}), since the remaining term is an even function of $\vc{q}$ according to our previous discussion,
\begin{equation}\label{eq:evenness}
\begin{split}
&\mathrm{Tr}\,H^\downarrow(-\vc{q})-\frac{1}{\beta}\sum_a\ln(1+e^{-\beta E_a(-\vc{q})})  = \mathrm{Tr}\,H^\uparrow(\vc{q})-\frac{1}{\beta}\sum_a\ln(1+e^{-\beta (-E_a(\vc{q}))}) \\ &= \mathrm{Tr}\,H^\uparrow(\vc{q}) - \mathrm{Tr}H_{\rm BdG}(\vc{q})  -\frac{1}{\beta}\sum_a\ln(1+e^{-\beta E_a(\vc{q})}) = \mathrm{Tr}\,H^\downarrow(\vc{q}) -\frac{1}{\beta}\sum_a\ln(1+e^{-\beta E_a(\vc{q})})\,.
\end{split}\end{equation}
This implies that $\vc{q} = 0$ is a stationary point of the grand potential, a candidate ground state. In order to check the stability of this stationary point it is necessary to calculate the second derivative with respect to $\vc{q}$. This amounts to the evaluation of the \textit{superfluid weight} $D_{\rm s}$. The superfluid weight is defined as the second derivative of the \textit{free energy} $F(T,N,\Delta_{\vc{i}}(\vc{q}),\vc{q}) = \Omega(T,\mu(\vc{q}),\Delta_{\vc{i}}(\vc{q}),\vc{q})+N\mu(\vc{q})$ since the total particle number is constant, 

\begin{equation}
[D_{\rm s}]_{i,j} = \frac{1}{V\hbar^2}\frac{\partial^2 F}{\partial q_i\partial  q_j} =  \frac{1}{V\hbar^2 }\frac{\partial^2 (\Omega+\mu N)}{\partial q_i\partial q_j}\,,
\end{equation}
where $V$ is the system  volume (in two dimensions the area $A$ is used instead).
In general the superfluid weight in an anisotropic system depends on the direction and is thus a tensor. As emphasised in our notation the quasiparticle energies and thus the grand potential depend on $\vc{q}$ both through $\mu(\vc{q})$ and $\Delta(\vc{q})$ and directly through the diagonal terms of the BdG Hamiltonian. It will be now shown that in fact it is not necessary to know the dependence of $\mu(\vc{q}),\Delta(\vc{q})$ to calculate the superfluid density and only their $\vc{q}= 0$ value is needed. 
This result has been proved in Ref.~\ncit{Taylor:2006}{2}, but we repeat the proof here with some 	modifications. By a suitable gauge transformation $\hat{c}_{\vc{i}\sigma} \to e^{i\phi_{\vc{i}}(\vc{q})}\hat{c}_{\vc{i}\sigma}$ it is possible to take $\Delta_{\vc{i}}(\vc{q})$ real and at the same time preserving the property~(\ref{eq:sym1}). This gauge will be used in the following. Taking the total derivative of the free energy with respect to $\vc{q}$ gives
\begin{equation}
\frac{\partial F}{\partial q_i} = \sum_{\vc{i}}\left.\frac{\partial \Omega}{\partial \Delta_{\vc{i}}}\right|_{\mu,\vc{q}} \frac{\partial \Delta_{\vc{i}}}{\partial q_i} + \left(\left.\frac{\partial \Omega}{\partial \mu}\right|_{\Delta_{\vc{i}},\vc{q}} + N\right)\frac{\partial \mu}{\partial q_i} + \left.\frac{\partial \Omega}{\partial q_i}\right|_{\Delta_{\vc{i}},\mu} = \left.\frac{\partial\Omega}{\partial q_i}\right|_{\Delta_{\vc{i}},\mu}\,.
\end{equation}
This result is valid for arbitrary $\vc{q}$ and it relies on the fact that $\mu$ and $\Delta_{\vc{i}}$
satisfy Eqs.~(\ref{eq:self_delta})-(\ref{eq:self_number}), i.e. they are self-consistent solutions. As a consequence a stationary point of the free energy corresponds to the condition $\left.\partial\Omega/\partial q_i\right|_{\Delta_{\vc{i}},\mu}=0$ on the grand potential which is satisfied for $\vc{q} = 0$ as we have seen above. 
From Eqs.~(\ref{eq:sym1})-(\ref{eq:sym3}) and the fact that $\Delta_{\vc{i}}(\vc{q})$ is real, one has
$\partial \Delta_{\vc{i}}/\partial q_i = \partial \mu / \partial q_i = 0$ at $\vc{q}=0$. This last result is useful when taking one more derivative of the free energy and evaluating it at equilibrium
\begin{equation}\label{eq:res1}
\left.\frac{\partial^2 F}{\partial q_j\partial q_i}\right|_{\vc{q}=0} = \sum_{\vc{i}}\left.\frac{\partial^2 \Omega}{\partial \Delta_{\vc{i}} \partial q_i} \frac{\partial \Delta_{\vc{i}}}{\partial q_j}\right|_{\vc{q}=0} +\left.\frac{\partial^2 \Omega}{\partial \mu \partial q_i} \frac{\partial \mu}{\partial q_j}\right|_{\vc{q}=0} + \left.\frac{\partial \Omega}{\partial q_j\partial q_i}\right|_{\vc{q}=0} = \left.\frac{\partial \Omega}{\partial q_j\partial q_i}\right|_{\vc{q}=0}\,.
\end{equation}
In the above equation the partial derivatives of the grand potential with respect to $\Delta_{\vc{i}},\mu,q_i$ are always taken keeping the other quantities fixed (introducing a notation to specify this would be quite cumbersome). The derivatives of the free energy with respect to $q_i$ instead are total derivatives in the sense that the dependence of $\mu(\vc{q}),\Delta(\vc{q})$ on $\vc{q}$ needs to be taken into account and the corresponding derivatives performed. 
The final result is then
\begin{equation}
[D_{\rm s}]_{i,j} = \left.\frac{1}{V\hbar^2} \frac{\partial^2 \Omega}{\partial q_i\partial q_j}\right|_{\Delta_{\vc{i}},\mu,\vc{q}=0}\,.
\end{equation}
The important point is that the above derivatives are taken for $\Delta_{\vc{i}},\mu$ constant even if they are themselves functions of $\vc{q}$.  Only the value of the order parameters at $\vc{q} = 0$ is needed to calculate the superfluid weight. This is unsurprising since the superfluid weight is a linear response coefficient and therefore it depends only on the properties of the ground state. 

\subsection*{Supplementary Note 2: Wannier functions and tight-binding Hamiltonian}
Consider the Schr\"odinger equation with a periodic potential $V(\vc{r})= V(\vc{r}+\vc{a}_i)$, with $\vc{a}_i$ the fundamental vectors that define the Bravais lattice,
\begin{equation}
H\psi(\vc{r}) = \left(-\frac{\hbar^2}{2m}\vc{\nabla}^2 + V(\vc{r})\right)\psi({\vc{r}}) = \varepsilon\psi(\vc{r})\,.
\end{equation}
Bloch theorem states that the solutions can be labelled by a band index $n$ and a quasimomentum index $\vc{k}$ and take the form\ucit{Grosso_Book}{3}
\begin{equation}
\psi_{n\vc{k}}(\vc{r}) = e^{i\vc{k}\cdot\vc{r}}g_{n\vc{k}}(\vc{r})\,, \quad \text{with eigenvalue} \quad \varepsilon_{n\vc{k}}\,. 
\end{equation}
The functions $g_{n\vc{k}}(\vc{r})$, called periodic Bloch functions (Bloch functions for brevity), are periodic in the real space coordinate $g_{n\vc{k}}(\vc{r}) = g_{n\vc{k}}(\vc{r}+\vc{a}_i)$. The eigenvalues are periodic in the quasimomentum $\varepsilon_{n\vc{k}} = \varepsilon_{n(\vc{k}+\vc{d}_i)}$, with $\vc{d}_i$ the fundamental vectors of the reciprocal lattice that are defined by $\vc{a}_i\cdot \vc{d}_j = 2\pi \delta_{ij}$. We adopt the convention that the  periodic Bloch functions are normalized $\int_{\Omega} d^3\vc{r}\,|g_{n\vc{k}}(\vc{r})|^2 = 1$, where the symbol $\int_{\Omega}d^3\vc{r}$ denotes the integration over one unit cell. The band index $n$ runs from 1 to $+\infty$ since there are an infinite number of bands in the continuum. Consider now a subset $\mathcal{S}$ of relevant bands, also called \textit{composite band}, separated from other bands below and above by sufficiently large bands gaps. 
It is possible to construct a particular convenient basis of localized functions that span exactly the same subspace spanned by the Bloch plane waves $\psi_{n\vc{k}}$ relative to the composite band. These are called Wannier functions and are defined by
\begin{equation}\label{eq:definition}
\begin{split}
w_\alpha(\vc{r}-\vc{r}_{\vc{i}}) &= \frac{V_\Omega}{(2\pi)^3}\int d^3\vc{k}\,e^{-i\vc{k}\cdot\vc{r}_{\vc{i}}}\sum_{n\in \mathcal{S}}[U_{\vc{k}}]_{\alpha,n}\psi_{n\vc{k}}(\vc{r}) \\
&= \frac{V_\Omega}{(2\pi)^3}\int d^3\,\vc{k}e^{i\vc{k}\cdot(\vc{r}-\vc{r}_{\vc{i}})}\sum_{n\in \mathcal{S}}[U_{\vc{k}}]_{\alpha,n}g_{n\vc{k}}(\vc{r}) \\
&= \frac{V_\Omega}{(2\pi)^3}\int d^3\,\vc{k}e^{i\vc{k}\cdot(\vc{r}-\vc{r}_{\vc{i}})}\sum_{n\in \mathcal{S}}[U_{\vc{k}}]_{\alpha,n}g_{n\vc{k}}(\vc{r}-\vc{r}_{\vc{i}})\,.
\end{split}
\end{equation}
$V_{\Omega}$ is the volume of the unit cell (in two dimension the area $A_\Omega$ of the unit cell is used instead) and the vector $\vc{r}_{\vc{i}}= i_x\vc{a}_1+i_y\vc{a}_2+i_z\vc{a}_3$, labelled by a triplet of integers $\vc{i} = (i_x,i_y,i_z)^T$, is a generic vector of the lattice. In the last line of the above equation we have made use of the fact that $g_{n\vc{k}}(\vc{r}) = g_{n\vc{k}}(\vc{r}-\vc{r}_{\vc{i}})$ for arbitrary $\vc{i}$. 
The matrix $U_{\vc{k}}$ is a $\vc{k}$-dependent unitary matrix that has to be chosen in order to obtain properly localized Wannier functions.

The Wannier functions $w_\alpha(\vc{r}-\vc{r}_{i})$ in Eq.~(\ref{eq:definition}) are obtained by the Fourier transform of the Bloch functions $\psi_{n\vc{k}}(\vc{r})$. The Fourier transform is a unitary operator, therefore the orthonormality of the Bloch functions is inherited by the Wannier functions
\begin{equation}
\int d^3\vc{r}\, \psi^*_{n\vc{k}}(\vc{r})\psi_{n'\vc{k}'}(\vc{r}) = \frac{(2\pi)^3}{V_{\Omega}} \delta_{n,n'}\delta(\vc{k}-\vc{k}') \Rightarrow \int d^3\vc{r}\, w^*_{\alpha}(\vc{r}-\vc{r}_{\vc{i}})w_{\beta}(\vc{r}-\vc{r}_{\vc{j}}) = \delta_{\alpha,\beta}\delta_{\vc{i},\vc{j}}\,.
\end{equation}
According to Ref.~\ncit{Brouder:2007}{4} if the Chern numbers (one number in 2D and three numbers in 3D) of the composite band are zero then it is possible to choose $U_{\vc{k}}$ is such a way that the Wannier functions are exponentially localized. Moreover there are practical methods to calculate the matrices $U_{\vc{k}}$ that guarantee that the Wannier functions are maximally localized, more precisely they minimize a suitable localization functional (see \hyperref[sec:bound_loc_func]{Supplementary Note 5}). The Wannier functions are useful since they are a basis of states for the subspace of the Hilbert space relative to the given composite band and are very simple since they are translated copies by displacements $\vc{r}_{\vc{i}}$ of the small set of functions $\{w_{\alpha}(\vc{r})\}$. The centers $\vc{r}_{\vc{i}\alpha}$ of the Wannier functions are defined   as
\begin{equation}
\vc{r}_{\vc{i}\alpha} = i_x\vc{a}_1+i_y\vc{a}_2+i_z\vc{a}_3 +\vc{b}_{\alpha}\,,\quad \text{with}\quad 
\vc{b}_{\alpha}=\int d^3\vc{r}\,|w_\alpha(\vc{r})|^2\vc{r}\,.
\end{equation}
We now consider the Hamiltonian in second quantized form in the basis of Wannier functions. The braket notation is used for the Wannier functions $\braket{\vc{r}}{\vc{i}\alpha} = w_{\alpha}(\vc{r}-\vc{r}_{\vc{i}})$. The second quantized quadratic Hamiltonian is $\mathcal{\hat H} = \int d^3\vc{r}\,\hat\psi^\dagger(\vc{r})H\hat\psi(\vc{r})$ with $\hat\psi(\vc{r})$ the fermionic field operator satisfying the usual anticommutation relations $\{\hat\psi(\vc{r}),\hat\psi^\dagger(\vc{r}')\} = \delta(\vc{r}-\vc{r}')$. We expand the field operator in the basis of Wannier functions
\begin{equation}
\hat \psi(\vc{r}) = \sum_{\vc{i}}\sum_{\alpha \in \mathcal{S}} w_{\alpha}(\vc{r}-\vc{r}_{\vc{i}})\hat{c}_{\vc{i}\alpha} +\sum_{\vc{i}} \sum_{\gamma \in \mathcal{\bar{S}}}\bar{w}_{\gamma}(\vc{r}-\vc{r}_{\vc{i}})\hat{d}_{\vc{i}\gamma}\,.
\end{equation}
The functions $\bar{w}_\gamma(\vc{r}-\vc{r}_{\vc{i}})$ are Wannier functions of the bands not included in the chosen composite band, and belong to the complementary set $\mathcal{\bar S}$.
Inserting this expansion into the expression for $\mathcal{\hat H}$ one obtains
\begin{equation}\label{eq:hopping_second_quantized}
\begin{split}
\mathcal{\hat H} = \int d^3\vc{r}\,\hat\psi^\dagger(\vc{r})H\hat{\psi}(\vc{r}) &= \sum_{\alpha,\beta \in \mathcal{S}}\sum_{\vc{i},\vc{j}} \hat{c}^\dagger_{\vc{i}\alpha} \bra{\vc{i}\alpha}H\ket{\vc{j}\beta}\hat{c}_{\vc{j}\beta} +  
\sum_{\gamma,\delta \in \mathcal{\bar S}}\sum_{\vc{i},\vc{j}} \hat{d}^\dagger_{\vc{i}\gamma} \bra{\vc{i}\gamma}H\ket{\vc{j}\delta}\hat{d}_{\vc{j}\delta}\\
& = \sum_{\alpha,\beta \in \mathcal{S}}\sum_{\vc{i},\vc{j}} \hat{c}^\dagger_{\vc{i}\alpha} K_{\vc{i}\alpha,\vc{j}\beta}\hat{c}_{\vc{j}\beta} +  
\sum_{\gamma,\delta \in \mathcal{\bar S}}\sum_{\vc{i},\vc{j}} \hat{d}^\dagger_{\vc{i}\gamma} \bar{K}_{\vc{i}\gamma,\vc{j}\delta}\hat{d}_{\vc{j}\delta}\,.
\end{split}
\end{equation}
The  key point in the above equation is that matrix elements of the form $\bra{\vc{i}\alpha}H\ket{\vc{j}\gamma}$ between a Wannier function belonging to the composite band  ($\alpha \in \mathcal{S}$) and a Wannier function in the complementary set ($\gamma \in \bar{\mathcal{S}}$) vanish, the reason being that they are built from eigenfunctions of the Hamiltonian that belong to orthogonal subspaces.
On the other hand off-diagonal matrix elements $\bra{\vc{i}\alpha}H\ket{\vc{j}\beta}$ with $\alpha,\beta \in \mathcal{S}$ are in general nonzero due to the fact that the Wannier functions are not eigenstates of the Hamiltonian. In particular there is a coupling between Wannier functions with $\alpha \neq \beta$ due to the mixing given by the unitary matrix $U_{\vc{k}}$ in the definition~(\ref{eq:definition}). From the definition~(\ref{eq:hopping_second_quantized}) it is also evident that the hopping matrix elements are translationally invariant $K_{\vc{i}\alpha,\vc{j}\beta} = K_{\alpha,\beta}(\vc{i}-\vc{j})$. For exponentially localized Wannier function the matrix elements $K_{\alpha,\beta}(\vc{i}-\vc{j})$ decay rapidly with the distance $\vc{i}-\vc{j}$ between unit cells and it is usually a good approximation to retain only the matrix elements between a limited number of neighbouring Wannier functions. The result is the so-called tight-binding Hamiltonian\ucit{Grosso_Book}{3} which provides the physical picture of particles moving by discrete tunneling events between the sites of the lattice.

Finally Eqs.~(28)-(29) in the main text can be obtained by expanding the field operators in plane waves $\hat{c}_{\vc{i}\alpha\sigma} = \frac{1}{\sqrt{N_{\rm c}}}\sum_{\vc{k}}e^{i\vc{k}\cdot\vc{r}_{\vc{i}\alpha}}\hat{c}_{\alpha\vc{k}\sigma}$ ($N_{\rm c}$ is the number of unit cells)
\begin{equation}
\begin{split}
&\sum_{\alpha,\beta \in \mathcal{S}}\sum_{\vc{i},\vc{j}} \hat{c}^\dagger_{\vc{i}\alpha\sigma} K_{\vc{i}\alpha,\vc{j}\beta}^\sigma\hat{c}_{\vc{j}\beta\sigma} = \frac{1}{N_{\rm c}}\sum_{\alpha,\beta \in \mathcal{S}}\sum_{\vc{i},\vc{j}}\bigg(\sum_{\vc{k}_1}e^{-i\vc{k}_1\cdot\vc{r}_{\vc{i}\alpha}}\hat{c}^\dagger_{\alpha\vc{k}_1\sigma}\bigg)
K_{\vc{i}\alpha,\vc{j}\beta}^\sigma
\bigg(\sum_{\vc{k}_2}e^{i\vc{k}_2\cdot\vc{r}_{\vc{j}\beta}}
\hat{c}_{\beta\vc{k}_2\sigma}\bigg) \\ &= \sum_{\alpha,\beta \in \mathcal{S}}
\sum_{\vc{k}_1,\vc{k}_2}e^{-i\vc{k}_1\cdot\vc{b}_\alpha}\hat{c}^\dagger_{\alpha\vc{k}_1\sigma} e^{i\vc{k}_2\cdot\vc{b}_\beta}\hat{c}_{\beta\vc{k}_2\sigma}
\bigg(\frac{1}{N_{\rm c}}\sum_{\vc{j}}e^{i(\vc{k}_2-\vc{k}_1)\cdot\vc{r}_{\vc{j}}}\bigg)
\sum_{\vc{i}-\vc{j}}e^{-i\vc{k}_1\cdot\vc{r}_{\vc{i}-\vc{j}}}K_{\alpha,\beta}^\sigma(\vc{i}-\vc{j})\\
&=\sum_{\alpha,\beta \in \mathcal{S}}\sum_{\vc{k}}\hat{c}^\dagger_{\alpha\vc{k}\sigma}\hat{c}_{\beta\vc{k}\sigma}
\bigg(\sum_{\vc{i}-\vc{j}}e^{-i\vc{k}\cdot(\vc{r}_{\vc{i}\alpha}-\vc{r}_{\vc{j}\beta})}K_{\vc{i}\alpha,\vc{j}\beta}^\sigma\bigg) = \sum_{\alpha,\beta \in \mathcal{S}}\sum_{\vc{k}}\hat{c}^\dagger_{\alpha\vc{k}\sigma}[\widetilde{K}^\sigma(\vc{k})]_{\alpha,\beta}\hat{c}_{\beta\vc{k}\sigma}\,.
\end{split}
\end{equation}

\subsection*{Supplementary Note 3: Superfluid weight in a multiband system at finite temperature}
\label{sec:general_formula}

From now on we consider a translational invariant Hamiltonian and we use the notation employed in the main text. The symmetry constrain in Eq.~(\ref{eq:sym_diag}) implies that  for $\vc{q}=0$ the matrices $E_{\vc{k}}(\vc{q})$ and $\mathcal{W}_{\vc{k}}(\vc{q})$ have the form shown in Eqs.~(8) and (9) in the main text.
It is convenient to rewrite the grand potential in Eq.~(\ref{eq:grand_potential_2}) in a slightly different form which can be derived in a way similar to Eq.~(\ref{eq:evenness})
\begin{equation}
\Omega(T,\mu,\Delta,\vc{q}) = \sum_{\vc{k}}\frac{1}{2}\big(\mathrm{Tr}\,H_{\vc{k}}^\uparrow(\vc{q})+\mathrm{Tr}\,H_{\vc{k}}^\uparrow(-\vc{q})\big) -\frac{1}{2\beta}\sum_{\vc{k}}\mathrm{Tr}\left[\ln 2\big(1+\cosh\beta H_{\vc{k}}(\vc{q})\big)\right]+N_c\frac{\mathrm{Tr}|\Delta(\vc{q})|^2}{U}\,.
\end{equation}
According to the results in the previous section the last term $\mathrm{Tr}|\Delta(\vc{q})|^2/U = \sum_\alpha |\Delta_\alpha(\vc{q})|^2/U$ can be dropped since it does not contribute to the superfluid density. Also the first term $\frac{1}{2}\big(\mathrm{Tr}\,H^\uparrow(\vc{q})+\mathrm{Tr}\,H^\uparrow(-\vc{q})\big) = \frac{1}{2}\sum_{\vc{k}} \big(\mathrm{Tr}\left[\varepsilon_{\vc{k}-\vc{q}}-\mu\bm{1}\right]+
\mathrm{Tr}\left[\varepsilon_{\vc{k}+\vc{q}}-\mu\bm{1}\right]\big)$ does not contribute in the thermodynamic limit since $\varepsilon_{\vc{k}}$ is a periodic function on the Brillouin zone. However we keep this term for reasons that will be clear in the following. Note that in the low temperature limit $\beta \to +\infty$ the second term reduces to $-\frac{1}{2}\mathrm{Tr}\left[|H_{\vc{k}}(\vc{q})|\right]$ in agreement with Eq.~(10) in the main text.

The first derivative of the grand potential with respect to $\vc{q}$ is the current density
\begin{equation}\label{eq:current}
\begin{split}
\vc{J}(\vc{q}) &= \frac{1}{V\hbar } \left.\frac{\partial \Omega}{\partial \vc{q}}\right|_{\mu,\Delta} = \frac{1}{V\hbar}\sum_{\vc{k}}\bigg(\frac{1}{2}\left(\mathrm{Tr}[\partial_{\vc{k}}\varepsilon_{\vc{k}+\vc{q}}]
-\mathrm{Tr}[\partial_{\vc{k}}\varepsilon_{\vc{k}-\vc{q}}]\right) - \frac{1}{2}\mathrm{Tr}\left[\mathrm{tanh}\left(\frac{\beta E_{\vc{k}}(\vc{q})}{2}\right)\partial_{\vc{q}}E_{\vc{k}}(\vc{q})
\right]\bigg)\,.
\end{split}
\end{equation}
The matrix elements of the diagonal matrix $\partial_{\vc{q}}E_{\vc{k}}(\vc{q})$ can be calculated using the Hellman-Feynman theorem and the expression for $H_{\vc{k}}(\vc{q})$ in Eq.~(6)  in the main text
\begin{gather}\label{eq:hellman-feynman}
\left[\partial_{\vc{q}} E_{\vc{k}}(\vc{q})\right]_{a,a} = \left[\mathcal{W}_{\vc{k}}^\dagger(\vc{q})\, \partial_{\vc{q}}H_{\vc{k}}(\vc{q})\,
\mathcal{W}_{\vc{k}}(\vc{q})\right]_{a,a}\,,
\qquad\text{with}\qquad 
\partial_{\vc{q}}H_{\vc{k}}(\vc{q}) = -
\begin{pmatrix}
\partial_{\vc{k}}\varepsilon_{\vc{k}-\vc{q}} & \partial_{\vc{q}}\mathcal{D}_{\vc{k}}(\vc{q}) \\
-\partial_{\vc{q}}\mathcal{D}_{\vc{k}}(-\vc{q}) & \partial_{\vc{k}}\varepsilon_{\vc{k}+\vc{q}}
\end{pmatrix}\,.
\end{gather}
We have defined $\mathcal{D}_{\vc{k}}(\vc{q}) = - \mathcal{G}_{\vc{k}-\vc{q}}^\dagger\Delta\mathcal{G}_{\vc{k}+\vc{q}} = \mathcal{D}^{\dagger}_{\vc{k}}(-\vc{q})$ in the above equation.
Notice that the Hellman-Feynman theorem holds only for the diagonal components $a=b$ of the matrices on the left and right hand side as emphasised by the square bracket notation $\left[\,\cdot\,\right]_{a,b=a}$. The matrix $\mathrm{tanh}(E_{\vc{k}}(\vc{q}))$ is itself diagonal and only the diagonal matrix elements in Eq.~(\ref{eq:hellman-feynman}) are needed in the evaluation of the trace in Eq.~(\ref{eq:current}).
It is easy to check that the current is an odd function using the identities in Eq.~(\ref{eq:sym_diag}). Again one can distinguish a conventional  current contribution and a contribution that comes from the off-diagonal terms in $\partial_{\vc{q}}H_{\vc{k}}(\vc{q})$ as in the zero temperature case discussed in the main text. Indeed the two contributions are separately gauge invariant as explained below. 

When taking an additional derivative with respect to $\vc{q}$ there will be terms proportional to the second derivative of the BdG Hamiltonian $\partial_{q_i}\partial_{q_j}H_{\vc{k}}(\vc{q})$ and first derivatives of the unitary matrix $\partial_{\vc{q}}\mathcal{W}_{\vc{k}}(\vc{q})$. The following identity is useful in order to express the off-diagonal matrix elements 
of $\mathcal{W}_{\vc{k}}^\dagger(\vc{q})\partial_{\vc{q}}\mathcal{W}_{\vc{k}}(\vc{q}) = -\big(\partial_{\vc{q}}\mathcal{W}_{\vc{k}}^\dagger(\vc{q})\big)\mathcal{W}_{\vc{k}}(\vc{q})$ only in terms of the ground state solution
\begin{equation}\label{eq:off-diagonal_derivative}
\left[\mathcal{W}_{\vc{k}}^\dagger(\vc{q})\partial_{\vc{q}}\mathcal{W}_{\vc{k}}(\vc{q})\right]_{a,b} = -\frac{\left[\mathcal{W}_{\vc{k}}^\dagger(\vc{q})\, \partial_{\vc{q}}H_{\vc{k}}(\vc{q})\,
\mathcal{W}_{\vc{k}}(\vc{q})\right]_{a,b}}{[E_{\vc{k}}(\vc{q})]_{a,a}-[E_{\vc{k}}(\vc{q})]_{b,b}} \quad \text{for} \quad a\neq b\,.
\end{equation}
The second derivatives of the quasiparticle excitaton energies can therefore be written
as 
\begin{equation}\label{eq:double_der_E}
\begin{split}
&[\partial_{q_i}\partial_{q_j}E_{\vc{k}}(\vc{q})]_{a,a} = \left[\mathcal{W}_{\vc{k}}^\dagger(\vc{q})\, \partial_{q_i}\partial_{q_j}H_{\vc{k}}(\vc{q})\,
\mathcal{W}_{\vc{k}}(\vc{q})\right]_{a,a}\\
&+\sum_{\substack{b\\a\neq b}} 
\frac{1}{[E_{\vc{k}}(\vc{q})]_{a,a}-[E_{\vc{k}}(\vc{q})]_{b,b}}\left( 
\left[\mathcal{W}_{\vc{k}}^\dagger(\vc{q})\, \partial_{q_i}H_{\vc{k}}(\vc{q})\,
\mathcal{W}_{\vc{k}}(\vc{q})\right]_{a,b}
\left[\mathcal{W}_{\vc{k}}^\dagger(\vc{q})\, \partial_{q_j}H_{\vc{k}}(\vc{q})\,
\mathcal{W}_{\vc{k}}(\vc{q})\right]_{b,a}+(i\leftrightarrow j )\right)\,.
\end{split}
\end{equation}
Taking one more derivative of the current density and setting $\vc{q}=0$ leads to the superfluid weight
\begin{equation}\label{eq:superfluid_v1}
\begin{split}
[D_{\rm s}]_{i,j} = \frac{1}{V\hbar^2}\sum_{\vc{k}}\bigg\{\mathrm{Tr}[\partial_{k_i}\partial_{k_j}\varepsilon_{\vc{k}}] 
-\frac{1}{2}\mathrm{Tr}\left[\mathrm{tanh}\left(\frac{\beta E_{\vc{k}}}{2}\right)\partial_{q_i}\partial_{q_j}E_{\vc{k}}\right]-\frac{1}{2}\mathrm{Tr}\left[\frac{\beta}{2\cosh^2(\beta E_{\vc{k}}/2)}\partial_{q_i}E_{\vc{k}}\partial_{q_j}E_{\vc{k}}\right]
\bigg\}\,,
\end{split}
\end{equation}
where $E_{\vc{k}} = E_{\vc{k}}(\vc{q}=0)$, $\partial_{q_i}E_{\vc{k}}= \partial_{q_i}E_{\vc{k}}(\vc{q}=0)$ and 
$\partial_{q_i}\partial_{q_j}E_{\vc{k}} = \partial_{q_i}\partial_{q_j}E_{\vc{k}}(\vc{q}=0)$.
Eq.~(\ref{eq:superfluid_v1}) can be written in a more convenient form by introducing some definitions (see Eqs.~(8)-(9) in the main text)
\begin{gather}
\begin{split}
&N_{\vc{k},i} = \mathcal{W}_{\vc{k}}^\dagger(\vc{q}=0)\, \partial_{q_i}H_{\vc{k}}(\vc{q}=0)\,
\mathcal{W}_{\vc{k}}(\vc{q}=0) = 
\begin{pmatrix}
A_{\vc{k},i} & B_{\vc{k},i} \\ -B_{\vc{k},i} & A_{\vc{k},i}
\end{pmatrix} \\ 
&\text{with}\;\;
\begin{cases}
A_{\vc{k},i} = \mathcal{U}_{\vc{k}}^\dagger\partial_{k_i}\varepsilon_{\vc{k}} \mathcal{U}_{\vc{k}} + \mathcal{V}_{\vc{k}}^\dagger\partial_{k_i}\varepsilon_{\vc{k}} \mathcal{V}_{\vc{k}} +
\mathcal{U}_{\vc{k}}^\dagger \partial_{q_i}\mathcal{D}_{\vc{k}} \mathcal{V}_{\vc{k}} - \mathcal{V}_{\vc{k}}^\dagger \partial_{q_i}\mathcal{D}_{\vc{k}} \mathcal{U}_{\vc{k}} 
\\
B_{\vc{k},i} = \mathcal{U}_{\vc{k}}^\dagger \partial_{q_i}\mathcal{D}_{\vc{k}} \mathcal{U}_{\vc{k}} + \mathcal{V}_{\vc{k}}^\dagger \partial_{q_i}\mathcal{D}_{\vc{k}} \mathcal{V}_{\vc{k}} 
+ \mathcal{V}_{\vc{k}}^\dagger\partial_{k_i}\varepsilon_{\vc{k}} \mathcal{U}_{\vc{k}} - \mathcal{U}_{\vc{k}}^\dagger\partial_{k_i}\varepsilon_{\vc{k}} \mathcal{V}_{\vc{k}}
\end{cases}
\end{split}
\label{eq:aux_def1}
\\
[T_{\vc{k}}]_{a,b} = 
\begin{cases}
\left[\dfrac{\beta}{2\cosh^2(\beta E_{\vc{k}}/2)}\right]_{a,a} & \text{for}\quad a = b\,, \\[1.5em]
\dfrac{[\tanh(\beta E_{\vc{k}}/2)]_{a,a}-[\tanh(\beta E_{\vc{k}}/2)]_{b,b}}{[E_{\vc{k}}]_{a,a}-[E_{\vc{k}}]_{b,b}} & \text{for}\quad a \neq b\,.
\end{cases}\label{eq:aux_def2}
\end{gather}
The abbreviation  $\partial_{q_i}\mathcal{D}_{\vc{k}} = \partial_{q_i}\mathcal{D}_{\vc{k}}(\vc{q}=0)$ has just been employed and $\partial_{q_i}\partial_{q_j}\mathcal{D}_{\vc{k}} = \partial_{q_i}\partial_{q_j}\mathcal{D}_{\vc{k}}(\vc{q}=0)$ will be employed in the following.
The final result for the superfluid weight tensor contains three contributions
\begin{gather}\label{eq:superfluid_weight_final_1}
[D_{\rm s}]_{i,j} = \left.\frac{1}{V\hbar^2}\frac{\partial^2\Omega}{\partial q_i\partial q_j}\right|_{_{\mu,\Delta,\vc{q}=0}} = [D_{{\rm s},1}]_{i,j} 
+ [D_{{\rm s},2}]_{i,j}+ [D_{{\rm s},3}]_{i,j}\,, \qquad \text{with}\\
[D_{{\rm s},1}]_{i,j} =  \frac{2}{V\hbar^2}\sum_{\vc{k}}\mathrm{Tr}\left[\left(\mathcal{V}_{\vc{k}}\frac{1}{e^{-\beta E_{\vc{k}}^>}+1}\mathcal{V}_{\vc{k}}^\dagger +
\mathcal{U}_{\vc{k}}\frac{1}{e^{\beta E_{\vc{k}}^>}+1}\mathcal{U}_{\vc{k}}^\dagger \right)\partial_{k_i}\partial_{k_j}\varepsilon_{\vc{k}}\right]\,, \label{eq:superfluid_weight_final_2}\\ 
[D_{{\rm s},2}]_{i,j} = \frac{2}{V\hbar^2}\sum_{\vc{k}}\mathrm{Tr}\left[\left(\mathcal{U}_{\vc{k}}\mathcal{V}_{\vc{k}}^\dagger -\mathcal{U}_{\vc{k}}\frac{1}{e^{\beta E^>_{\vc{k}}}+1}\mathcal{V}^\dagger_{\vc{k}}-\mathcal{V}_{\vc{k}}\frac{1}{e^{\beta E^>_{\vc{k}}}+1}\mathcal{U}_{\vc{k}}^\dagger\right)\partial_{q_i}\partial_{q_j}\mathcal{D}_{\vc{k}}
\right]\,,
\label{eq:superfluid_weight_final_3}\\
[D_{{\rm s},3}]_{i,j} = -\frac{1}{2V\hbar^2}\sum_{\vc{k}} \sum_{a,b} [T_{\vc{k}}]_{a,b}[N_{\vc{k},i}]_{a,b}[N_{\vc{k},j}]_{b,a}\,.\label{eq:superfluid_weight_final_4}
\end{gather}
We have introduced the diagonal matrix $E^>_{\vc{k}} = \mathrm{diag}(E_{n\vc{k}})$, i.e. the upper diagonal block of $E_{\vc{k}}$ which contains the positive eigenvalues $E_{n\vc{k}}>0$ (see Eq.~(8) in the main text). The $\vc{k}$-dependent effective mass tensor $\left[\frac{1}{m_{{\rm eff},\vc{k}}}\right]_{i,j} = \frac{1}{\hbar^2}\frac{\partial^2\varepsilon_{\vc{k}}}{\partial_{k_i}\partial_{k_j}}$ enters the expression of $D_{{\rm s},1}$ which is the combination of the first term in Eq.~(\ref{eq:superfluid_v1}) and the terms proportional to $\frac{\partial^2\varepsilon_{\vc{k}}}{\partial_{k_i}\partial_{k_j}}$ in the second term in the same equation (thus our derivation holds also in the case of a finite system for which $\sum_{\vc{k}}\mathrm{Tr}[\varepsilon_{\vc{k}+\vc{q}}]$ is not a constant as a function of $\vc{q}$). The analogous quantity that enters $D_{{\rm s},2}$ is
\begin{equation}\label{eq:anomalous_mass}
\partial_{q_i}\partial_{q_j}\mathcal{D}_{\vc{k}} = 
(\partial_{k_i}\mathcal{G}_{\vc{k}}^\dagger)\Delta
(\partial_{k_j}\mathcal{G}_{\vc{k}})
+(\partial_{k_j}\mathcal{G}_{\vc{k}}^\dagger)\Delta
(\partial_{k_i}\mathcal{G}_{\vc{k}}) -
\left(\partial_{k_i}\partial_{k_j}\mathcal{G}_{\vc{k}}^\dagger\right)\Delta\mathcal{G}_{\vc{k}} -\mathcal{G}_{\vc{k}}^\dagger\Delta\left(\partial_{k_i}\partial_{k_j}\mathcal{G}_{\vc{k}}\right)\,.
\end{equation}
The above expression simplifies if the diagonal matrix $\Delta$ is proportional to the identity, resulting in ($\Delta$ is now a scalar, not a matrix)
\begin{equation}\label{eq:Berry_strange}
\partial_{q_i}\partial_{q_j}\mathcal{D}_{\vc{k}} = 2\Delta\left[\partial_{k_i}\mathcal{G}_{\vc{k}}^\dagger
\partial_{k_j}\mathcal{G}_{\vc{k}}
+\partial_{k_j}\mathcal{G}_{\vc{k}}^\dagger
\partial_{k_i}\mathcal{G}_{\vc{k}}\right]\,.
\end{equation}
In deriving the above equation we have used the identity $\partial_{k_i}\partial_{k_j}(\mathcal{G}_{\vc{k}}^\dagger\mathcal{G}_{\vc{k}}) = 0$, a consequence of the fact that $\mathcal{G}^\dagger_{\vc{k}}\mathcal{G}_{\vc{k}} = \bm{1}$. The component of the superfluid weight $D_{s,3}$ results from the combination of the second term in Eq.~(\ref{eq:double_der_E}) and the last term in Eq.~(\ref{eq:superfluid_v1}). It contains both a term which is possibly finite at zero temperature (see Eq.~(14) in the main text) and a negative term which in the single-band case corresponds to the normal fluid component.
$D_{{\rm s},3}$ must be present in order to preserve the invariance of the superfluid weight, an observable quantity, under gauge transformations. In this context the gauge transformations are related to an inherent ambiguity in the definition of the Bloch wavefunctions namely the possibility of applying a transformation of the form $\mathcal{G}_{\vc{k}} \to \mathcal{G}_{\vc{k}}\mathcal{A}_{\vc{k}}$ with $\mathcal{A}_{\vc{k}}$ a unitary matrix subject to the additional constrain
\begin{equation}\label{eq:gauge_constrain}
[\varepsilon_{\vc{k}}, \mathcal{A}_{\vc{k}}] = 0\,.
\end{equation}
This ensures that the form of the BdG Hamiltonian $H_{\vc{k}}(\vc{q})$ in Eq.~(6) in the main  text is preserved under the transformation. If the band dispersions are nondegenerate then $\mathcal{A}_{\vc{k}}$ is simply a diagonal matrix of phases $\mathcal{A}_{\vc{k}} = \mathrm{diag}(e^{i\phi_n(\vc{k})})$. $D_{{\rm s},1}$ is invariant under such a transformation since one has $\mathcal{V}_{\vc{k}}\to \mathcal{A}_{\vc{k}}^\dagger \mathcal{V}_{\vc{k}}$ and Eq.~(\ref{eq:gauge_constrain}) implies that $\mathcal{A}_{\vc{k}}$ commutes with the derivatives of $\varepsilon_{\vc{k}}$. This means that also the sum $D_{{\rm s},2}+D_{{\rm s},3}$ is gauge invariant. Below we show in a specific case that, by a suitable gauge transformation, it is possible to make the component $D_{{\rm s},3}$ vanish leaving only $D_{{\rm s},1}$ and $D_{{\rm s},2}$. But it is not clear if this is possible in general and it is convenient to allow some freedom in the choice of gauge.
The calculation of the superfluid weight in the multiband case requires only the solution of the ground state problem, namely the quantities $\mathcal{U}_{\vc{k}},\mathcal{V}_{\vc{k}},E_{n\vc{k}},\Delta_\alpha$,  and the only additional ingredients with respect to the single-band case are the Bloch functions and their derivatives $\partial_{k_i}\mathcal{G}_{\vc{k}}$ and $\partial_{k_i}\partial_{k_j}\mathcal{G}_{\vc{k}}$ that enter in $D_{s,2}$ and $D_{s,3}$. It is easy to check that in the zero temperature limit Eqs.~(\ref{eq:aux_def1})-(\ref{eq:superfluid_weight_final_4}) reduce to the zero temperature result in the main text (Eqs.~(12)-(14)). Eqs.~(\ref{eq:aux_def1})-(\ref{eq:superfluid_weight_final_4}) are very general and can be applied to a number of lattice models.

We also provide explicit expressions in terms of the matrices $\mathcal{U}_{\vc{k}}$ and $\mathcal{V}_{\vc{k}}$ and the quasiparticle energies $E_{n\vc{k}}$ of some correlation functions that are useful for the implementation of the self-consistency loop numerically or to check the self-consistency of an analytical solution. The first is the so-called anomalous Green function
\begin{equation}\label{eq:correlator_1}
F_{\vc{i}\alpha,\vc{j}\beta} = \langle {\hat c}_{\vc{i}\alpha\uparrow}{\hat c}_{\vc{j}\beta\downarrow} \rangle = 
\frac{1}{N_{\rm c}}\sum_{\vc{k}} e^{i\vc{k}\cdot (\vc{r}_{\vc{i}\alpha}-\vc{r}_{\vc{j}\beta})}\left[\mathcal{G}_{\vc{k}}\left(\mathcal{U}_{\vc{k}}\mathcal{V}_{\vc{k}}^\dagger -\mathcal{U}_{\vc{k}}\frac{1}{e^{\beta E^>_{\vc{k}}}+1}\mathcal{V}^\dagger_{\vc{k}}-\mathcal{V}_{\vc{k}}\frac{1}{e^{\beta E^>_{\vc{k}}}+1}\mathcal{U}_{\vc{k}}^\dagger\right)
\mathcal{G}_{\vc{k}}^\dagger\right]_{\alpha,\beta}\,.
\end{equation}
The unitary of $\mathcal{W}_{\vc{k}}(\vc{q})$ implies that $\mathcal{U}_{\vc{k}}\mathcal{V}_{\vc{k}}^\dagger = \mathcal{V}_{\vc{k}}\mathcal{U}_{\vc{k}}^\dagger$  which means that Eq.~(\ref{eq:correlator_1}) is an Hermitian matrix $F_{\vc{i}\alpha,\vc{j}\beta} = F_{\vc{j}\beta,\vc{i}\alpha}^*$ and, therefore, the pairing potential $\Delta_{\alpha} = U\langle {\hat c}_{\vc{i}\alpha\uparrow}{\hat c}_{\vc{i}\alpha\downarrow} \rangle = UF_{\vc{i}\alpha,\vc{i}\alpha}$ is real. Another correlation function of interest is the usual Green function $G$
\begin{equation}\label{eq:correlator_2}
\begin{split}
G_{\vc{i}\alpha,\vc{j}\beta} &= \langle \hat{c}^\dagger_{\vc{j}\beta\uparrow}\hat{c}_{\vc{i}\alpha\uparrow}\rangle = \langle \hat{c}^\dagger_{\vc{i}\alpha\downarrow}\hat{c}_{\vc{j}\beta\downarrow}\rangle 
= \frac{1}{N_{\rm c}}\sum_{\vc{k}}e^{-i\vc{k}\cdot (\vc{r}_{\vc{i}\alpha}-\vc{r}_{\vc{j}\beta})}\left[\mathcal{G}_{\vc{k}}\left(\mathcal{U}_{\vc{k}}\frac{1}{e^{\beta E^>_{\vc{k}} }+1}\mathcal{U}_{\vc{k}}^{\dagger}+\mathcal{V}_{\vc{k}}\frac{1}{e^{-\beta E^>_{\vc{k}} }+1}\mathcal{V}_{\vc{k}}^{\dagger}
\right)\mathcal{G}_{\vc{k}}^{\dagger}\right]_{\alpha,\beta}\,.
\end{split}
\end{equation}
The total particle density is immediately obtained from the Green's function  $n_{\vc{i}\alpha} = \langle \hat{c}^\dagger_{\vc{i}\alpha\uparrow}\hat{c}_{\vc{i}\alpha\uparrow}\rangle + \langle \hat{c}^\dagger_{\vc{i}\alpha\downarrow}\hat{c}_{\vc{i}\alpha\downarrow}\rangle  = 2G_{\vc{i}\alpha,\vc{i}\alpha}$.

%

\subsection*{Supplementary Note 4: Superfluid density of the time-reversal invariant Harper-Hubbard model}\label{sec:superfluidity_Harper}

We now specialize our result for the superfluid density in a generic multiband system to the case of the Harper model. The band dispersions $\varepsilon_{n\vc{k}}$ and periodic Bloch functions $g_{n\vc{k}}$ of the Harper model with commensurate flux per plaquette $n_\phi = 1/Q = 1/N_{\rm orb}$ are given by the Harper equation
\begin{equation}\label{eq:Harper_eq}
\begin{split}
e^{ik_ya}g_{n\vc{k}}(\alpha+1) + e^{-ik_ya}g_{n\vc{k}}(\alpha-1) + 2\cos\left(k_xa -2\pi\frac{\alpha}{Q}\right)g_{n\vc{k}}(\alpha) = -\frac{\varepsilon_{n\vc{k}}}{J}g_{n\vc{k}}(\alpha)\,,
\end{split}
\end{equation}
with periodic boundary conditions $g_{n\vc{k}}(\alpha) = g_{n\vc{k}}(\alpha+Q)$.
A useful  property of the Bloch functions that can be derived from the definition is
\begin{equation}\label{eq:Harper-property}
\varepsilon_{n\left(\vc{k}+\frac{2\pi}{Qa}\hat{\vc{k}}_x\right)} = \varepsilon_{n\vc{k}}, \quad  
g_{n\left(\vc{k}+\frac{2\pi}{Qa}\hat{\vc{k}}_x\right)}(\alpha) = g_{n\vc{k}}(\alpha-1)\qquad \text{with}\quad \hat{\vc{k}}_x = \begin{pmatrix} 1 \\ 0 \end{pmatrix}\,. 
\end{equation}
By means of the above identity it is shown below that a  constant gap function $\Delta_{\alpha} = \Delta$ is a self consistent solution of the gap equation, therefore in the following $\Delta$ is a positive real number and not a matrix as defined previously. This implies that  $\mathcal{G}_{\vc{k}}^\dagger \Delta \mathcal{G}_{\vc{k}} = \Delta\mathcal{G}_{\vc{k}}^\dagger \mathcal{G}_{\vc{k}} = \Delta\bm{1}$. In this case the BdG Hamiltonian $H_{\vc{k}}(\vc{q}=0)$ can be easily diagonalized and the solution reads
\begin{gather}
[\mathcal{U}_{\vc{k}}]_{n,n'} = u_{n\vc{k}}\delta_{n,n'} \quad \text{with} \quad u_{n\vc{k}} = \frac{1}{\sqrt{2}}\left(1+\frac{\varepsilon_{n\vc{k}}-\mu}{E_{n\vc{k}}}\right)^{\frac{1}{2}} \,,\label{eq:U_n}\\
[\mathcal{V}_{\vc{k}}]_{n,n'} = v_{n\vc{k}}\delta_{n,n'} \quad \text{with} \quad v_{n\vc{k}} = \frac{1}{\sqrt{2}}\left(1-\frac{\varepsilon_{n\vc{k}}-\mu}{E_{n\vc{k}}}\right)^{\frac{1}{2}} \,,\label{eq:V_n}\\
E_{n\vc{k}} = \sqrt{(\varepsilon_{n\vc{k}}-\mu)^2+\Delta^2}\,. \label{eq:quasiparticle} 
\end{gather}
A fundamental object for what follows is the projector operator $P_{n\vc{k}}$ defined in terms of its matrix elements
\begin{equation}
[P_{n\vc{k}}]_{\alpha,\beta} = \big[\mathcal{G}_{\vc{k}}\big]_{\alpha,n}\big[\mathcal{G}_{\vc{k}}^\dagger\big]_{n,\beta}\,.
\end{equation}
This is a projection operator since it is positive $P_{n\vc{k}} >0$ and idempotent $P_{n\vc{k}}^2 = P_{n\vc{k}}$. The above definition is immediately generalized to the projection operator onto a given set of bands simply by summing over the corresponding subset of values of $n$.
Note that Eq.~(\ref{eq:Harper-property}) imply that 
\begin{equation}
\begin{split}
&\sum_m\left[P_{n\left(\vc{k}+\frac{2\pi m}{Qa}\hat{\vc{k}}_x\right)}\right]_{\alpha,\alpha} = \sum_m g_{n\left(\vc{k}+\frac{2\pi m}{Qa}\hat{\vc{k}}_x\right)}(\alpha)g^*_{n\left(\vc{k}+\frac{2\pi m}{Qa}\hat{\vc{k}}_x\right)}(\alpha) = \sum_m |g_{n\vc{k}}(\alpha-m)|^2 = 1\,.  
\end{split}
\end{equation}
Thus the gap equation at zero temperature reads
\begin{equation}\label{eq:gap_eq}
\begin{split}
\Delta &= \Delta_{\alpha} = U F_{\vc{i}\alpha,\vc{i}\alpha} = \frac{U}{N_{\rm c}}\sum_{\vc{k},n} u_{n\vc{k}}v_{n\vc{k}}\tanh\frac{\beta E_{n\vc{k}}}{2}[P_{n\vc{k}}]_{\alpha,\alpha} \\&=\frac{U}{N_{\rm c}}\sum_{n}\sum_{\vc{k}\in \,\text{red.B.Z.}} \frac{\Delta}{2E_{n\vc{k}}}\tanh\frac{\beta E_{n\vc{k}}}{2}\sum_{m}\left[P_{n\left(\vc{k}+\frac{2\pi m}{Qa}\hat{\vc{k}}_x\right)}\right]_{\alpha,\alpha} \\
&= \frac{U}{N_{\rm c}}\sum_{n}\sum_{\vc{k}\in \,\text{red.B.Z.}} \frac{\Delta}{2E_{n\vc{k}}}\tanh\frac{\beta E_{n\vc{k}}}{2}\,,
\end{split}
\end{equation}
In the second and third line of the above equation the vector $\vc{k}$ takes values on the reduced Brillouin zone $k_x,k_y \in [-\pi/(Qa),\pi/(Qa)]$ since the degeneracy due to Eq.~(\ref{eq:Harper-property}) has been taken care of by the summation over $m$ in the second line.
Analogously the average number of particles per lattice site $n_{\rm p} = n_{\vc{i}\alpha}$  is fixed by the equation
\begin{equation}
\frac{n_{\rm p}}{2} = \frac{1}{N_c}\sum_n\sum_{\vc{k}\in \,\text{red.B.Z.}}
 \frac{1}{2}\left(1 - \frac{\varepsilon_{n\vc{k}}-\mu}{E_{n\vc{k}}}\tanh\frac{\beta E_{n\vc{k}}}{2}\right)\,.\label{eq:mu_eq}
\end{equation}
If a solution of the coupled equations~(\ref{eq:gap_eq})-(\ref{eq:mu_eq}) is found then our initial choice of a constant pairing potential is self-consistent and this is in general the case since we have two equations with two unknowns.

Up to now all of our calculations have not involved any approximations other than the mean-field one. We now use some approximate results for the eigenfunctions and eigenvalues of the Harper model as detailed in Ref.~\ncit{Harper:2014}{5}. It has been shown that for $Q \gg 1$ the first lowest laying bands are separated by a gap which is an algebraic function of $Q$, while the bandwidth of each band is exponentially suppressed ($\mathrm{max}_{\vc{k}}\varepsilon_{n\vc{k}}  -\mathrm{min}_{\vc{k}}\varepsilon_{n\vc{k}}\propto e^{-\alpha Q}$ with $\alpha \approx 1$). This corresponds to the the fact that the Harper model bands approach the perfectly flat Landau levels in the continuum limit.
We call $\varepsilon_{\bar{n}} \approx \varepsilon_{\bar{n}\vc{k}} $ the average energy of the $\bar{n}$-th band and choose a value of the interaction strength $U$ such that Eq.~(15) in the main text is satisfied, meaning that interband effects and effects due to a nonflat band dispersion are neglected. As a consequence the summation over the wavevector $\vc{k}$ produces $N_{\rm c}/Q = N_{\rm c}n_{\phi}$ identical terms in Eq.~(\ref{eq:gap_eq})-(\ref{eq:mu_eq}). We write $n_{\rm p}/(2n_\phi) = \bar{n}+\nu$ with $\bar{n}$ the integer number of the highest (partially or completely) occupied band of the Harper model and with $0 < \nu < 1$ the filling of factor of the same band. The  filling factor $\nu$ is spin resolved and is the same for up and down spins. The bands are numbered starting from $n = 0$ up to $n = Q-1$ to conform with the usual Landau level notation. In the flat band approximation $\varepsilon_{\bar{n}\vc{k}}\approx \varepsilon_{\bar{n}}$, one has from Eq.~(\ref{eq:quasiparticle}) that $E_{\bar{n}\vc{k}} \approx E_{\bar{n}}$ and since we expect $U \sim E_{\bar{n}} \ll E_{n\vc{k}}$ for $n \neq \bar{n}$, only the $n = \bar{n}$ term is relevant in Eq.~(\ref{eq:gap_eq}). By cancelling $\Delta$ on both sides one obtains the equation for $E_{\bar{n}}$
\begin{equation}\label{eq:Ebar}
E_{\bar{n}} = \frac{Un_{\phi}}{2}\tanh\frac{\beta E_{\bar{n}}}{2}\,.
\end{equation}
This equation has nonzero solution only for ${\beta Un_{\phi}}{/4} > 1$. Indeed $k_{\rm B}T_{\rm c}  = Un_\phi/4$ is the mean-field critical temperature at half filling (see below). Close to the critical temperature it is possible to use the approximation $\tanh x\approx x-\frac{x^3}{3}$ and derive $E_{\bar{n}}(T\lesssim T_{\rm c}) \approx 2\sqrt{3}k_B T\sqrt{1-T/T_{\rm c}}$.
Using Eq.~(\ref{eq:mu_eq}) as well it is easy to obtain the approximate self consistent solution at finite temperature
\begin{gather}
\label{eq:En}
E_{n\vc{k}} =
\begin{cases}
E_{\bar{n}}, & n =\bar{n}\,\\
|\varepsilon_{n\vc{k}} - \mu|, & n \neq\bar{n}\,,
\end{cases}\\
\mu = \varepsilon_{\bar{n}} + Un_\phi \left(\nu -\frac{1}{2}\right)\,, \label{eq:mu_eq_approx}\\
\Delta =\frac{Un_\phi}{2} \sqrt{\left(\frac{2E_{\bar{n}}}{Un_\phi}\right)^2-\left(2\nu -1\right)^2}\,.\label{eq:delta_eq}
\end{gather}
If the argument in the square root of Eq.~(\ref{eq:delta_eq}) is negative then $\Delta = 0$ even if Eq.~(\ref{eq:Ebar}) has a solution. This can happen away from half-filling. Instead at half filling $\Delta = E_{\bar{n}}$. 
The zero temperature limit of the above solution is given by Eq.~(16)-(19) in the main text. Since the Harper model has been recently implemented in ultracold gases we test the validity of the isolated flat-band approximation (Eqs.~(\ref{eq:Ebar})-(\ref{eq:delta_eq})) against the full self-consistent solution of Eq.~(\ref{eq:gap_eq})-(\ref{eq:mu_eq}) in Supplementary Figure~\ref{fig:Harper_Delta}. For reasonable (not too small values) of $n_\phi$ and $U$ the flat-band limit is indeed realized to a good approximation.

In the derivation of the approximate self-consistent solution~(\ref{eq:Ebar})-(\ref{eq:delta_eq}) we have used the property~(\ref{eq:Harper-property}) specific of the Harper model. However it is generally valid in the case where $\Delta_\alpha = \Delta$ with $\Delta$ a real scalar. Indeed simply using this assumptions, which leads to  Eqs.~(\ref{eq:U_n})-(\ref{eq:quasiparticle}), and the condition~(15) in the main text one obtains the two coupled self-consistency equations
\begin{gather}
\Delta = \Delta_\alpha = U\frac{\Delta}{2E_{\bar{n}}}
\tanh\left(\frac{\beta E_{\bar{n}}}{2}\right)
\left(\frac{1}{N_{\rm c}}\sum_{\vc{k}}[P_{\bar{n}\vc{k}}]_{\alpha,\alpha}\right)\,,\\
\frac{n_{\rm p}}{2} = \frac{n_{\vc{i}\alpha}}{2} =
\left[\bar{n}+\frac{1}{2}\left(1-\frac{\varepsilon_{\bar{n}}-\mu}{E_{\bar{n}}}\right)\tanh\left(\frac{\beta E_{\bar{n}}}{2}\right)\right]
\left(\frac{1}{N_{\rm c}}\sum_{\vc{k}}[P_{\bar{n}\vc{k}}]_{\alpha,\alpha}\right)\,.
\end{gather}
The projector on a single band is normalized according to $\mathrm{Tr}\,P_{\bar{n}\vc{k}} = 1$ and this implies $N_{\rm c}^{-1}\sum_{\vc{k}}[P_{\bar{n}\vc{k}}]_{\alpha,\alpha} = N_{\rm orb}^{-1}$. The result in Eqs.~(\ref{eq:Ebar})-(\ref{eq:delta_eq}) is thus recovered with $n_{\phi} = N_{\rm orb}^{-1}$.

We are now in position to evaluate the various contributions to the superfluid weight. The diagonal superfluid weight is zero in the flat-band approximation. The other contributions can be evaluated
straightforwardly and the result is  
\begin{equation}\label{eq:superfluid_weight_flat}
[D_{\rm s}]_{i,j} = \frac{2}{\pi\hbar^2}\frac{\Delta^2(T)}{Un_\phi}\mathcal{M}_{ij}^{\rm R}\,,
\end{equation}
where the temperature enters only in the energy gap $\Delta(T)$.
This result reduces to Eq.~(20) in the main text at zero temperature. 
In deriving the above equation we have used the fact that $u_{n\vc{k}}v_{n\vc{k}} \approx 0$ for $n \neq \bar{n}$. We have also neglected terms with $n \neq \bar{n}$ in Eq.~(\ref{eq:superfluid_weight_final_4}) since $E_{\bar{n}} \ll E_{n \neq \bar{n}}$ according to Eq.~(\ref{eq:En}).
The matrix $\mathcal{M}_{ij}^{\rm R} = \mathrm{Re}(\mathcal{M}_{ij})$ is the real part of a Hermitian matrix defined as
\begin{equation}\label{eq:M-def}
\mathcal{M}_{ij} = \frac{2\pi}{A}\sum_{\vc{k}}\mathcal{B}_{ij}(\vc{k}) = \frac{1}{2\pi}\int_{\rm B.Z.} d^2\vc{k}\, \mathcal{B}_{ij}(\vc{k})= \frac{1}{2\pi}\int_{\rm B.Z.} d^2\vc{k}\,2\left(\mathrm{Tr}\left[(\partial_{k_i}\bar{\mathcal{G}}_{\vc{k}}^\dagger)( \partial_{k_j}\bar{\mathcal{G}}_{\vc{k}})\right] + \mathrm{Tr}\left[\bar{\mathcal{G}}_{\vc{k}}^\dagger (\partial_{k_i}\bar{\mathcal{G}}_{\vc{k}}) \bar{\mathcal{G}}_{\vc{k}}^\dagger (\partial_{k_j}\bar{\mathcal{G}}_{\vc{k}})\right]\right)\,.
\end{equation}
The integration is over the full Brillouin zone.
The first term in parenthesis in the above equation corresponds to $D_{{\rm s},2}$ while the other one corresponds to $D_{{\rm s},3}$. Crucially the matrix $\mathcal{M}$ is defined in terms of  the matrix $\bar{\mathcal{G}}_{\vc{k}}$ which is now a $Q \times 1$ column vector (in general a rectangular matrix), defined as the projection of $\mathcal{G}_{\vc{k}}$ on the Bloch wavefunctions corresponding to the $\bar{n}$-th band 
\begin{equation}\label{eq:Gbar-def}
[\bar{\mathcal{G}}_{\vc{k}}]_{\alpha,1} = [\mathcal{G}_{\vc{k}}]_{\alpha,\bar{n}}\,, \quad 
\bar{\mathcal{G}}^\dagger_{\vc{k}}\bar{\mathcal{G}}_{\vc{k}} = \bm{1}\,, \quad 
\bar{\mathcal{G}}_{\vc{k}}\bar{\mathcal{G}}_{\vc{k}}^\dagger = P_{\bar{n}\vc{k}}\,.
\end{equation}
It can be easily shown that if $\mathcal{\bar{G}}_{\vc{k}}$ is a square unitary then $\mathcal{M}$ is zero. In our context in fact a square unitary matrix is a pure gauge transformation which must not have observable consequences. The matrix $\mathcal{M}$ is also positive semidefinite $\mathcal{M}\geq 0$ since it can be written in the form
\begin{equation}
\mathcal{M}_{ij} = \frac{1}{2\pi}\int_{\rm B.Z.} d^2\vc{k}\,2\mathrm{Tr}\left[(\partial_{k_i}\bar{\mathcal{G}}_{\vc{k}}^\dagger)(\bm{1}-P_{\bar{n}\vc{k}})( \partial_{k_j}\bar{\mathcal{G}}_{\vc{k}})\right] \,.
\end{equation}
The invariant quantity $\mathcal{B}_{ij}(\vc{k})$ is called the quantum geometric tensor, its real part is the quantum metric, while the imaginary part of $\mathcal{B}_{ij}(\vc{k})$ is the well-known Berry curvature, and its integral over the Brillouin zone is the Chern number $\mathrm{Im}(\mathcal{M}_{ij}) = \mathcal{M}^{\rm I}_{ij} = \epsilon_{ij}C$, generalized to a set of bands (a composite band)\ucit{Brouder:2007}{4} ($\epsilon_{ij}= - \epsilon_{ji}$ is the Levi-Civita symbol). The second term in parenthesis in Eq.~(\ref{eq:M-def}) is real and does not show up in the Berry curvature. In can be easily verified that the quantum geometric tensor is an invariant under gauge transformations $\bar{\mathcal{G}}_{\vc{k}} \to \bar{\mathcal{G}}_{\vc{k}} \mathcal{A}_{\vc{k}}$ where $\mathcal{A}_{\vc{k}}$ is a unitary matrix with dimension equal to the number of columns of $\bar{\mathcal{G}}_{\vc{k}}$.

We now want to evaluate $\mathcal{M}_{ij}$ using a suitable approximation for the wavefunctions of the Harper model. Such an approximation can be obtained by taking the Harper equation~(\ref{eq:Harper_eq}) and replacing the cosine potential with its second order expansion around the minimum and furthermore replacing the discrete differences with derivatives. We obtain the equation of an ordinary harmonic oscillator for the Bloch wavefunction $\psi(\vc{r}_{\vc{i}\alpha}) = e^{ik_xi_xa}\psi(\alpha + Qi_y)$
\begin{equation}\label{eq:harmonic_oscil}
-\frac{1}{2}\psi''(\alpha)+\frac{1}{2}\omega^2\left(\alpha-\frac{Qk_xa}{2\pi}\right)^2\psi(\alpha) = \frac{1}{2}\left(\frac{\varepsilon}{J}+4\right)\psi(\alpha) \quad \text{with} \quad \omega = \frac{2\pi}{Q}\,.
\end{equation}
The variable $\alpha$ is now a continuous variable. We are neglecting effects due to the tunneling between the minima of the cosine potential in the Harper equation~(\ref{eq:Harper_eq}) in the main text. These corrections can be shown to be of the same order of the exponentially small bandwidth\ucit{Harper:2014}{5} which we neglect in our treatment. The above equation is precisely the equation that one obtains when solving the problem of a quantum mechanical particle in a magnetic field introduced by means of the Landau gauge (note the shift of the harmonic potential proportional to $k_x$). The solutions are the harmonic oscillator wavefunctions $\varphi_n(x)$ (Gaussian function times an Hermite polynomial) which corresponds to the different Landau levels. It is also useful to introduce annihilation and creation operators to manipulate such functions ($\hat{x}=x$ and $\hat{p} = -i\partial_x$)
\begin{gather}
\hat{a} = \sqrt{\frac{\omega}{2}}\hat{x}+\frac{i}{\sqrt{2\omega}}\hat{p}, \quad \hat{a}\varphi_n = \sqrt{n}\varphi_{n-1}\,,\\
\hat{a}^\dagger = \sqrt{\frac{\omega}{2}}\hat{x}-\frac{i}{\sqrt{2\omega}}\hat{p}, \quad \hat{a}^\dagger\varphi_n = \sqrt{n+1}\varphi_{n+1}\,.
\end{gather}
Eq.~(\ref{eq:harmonic_oscil}) can be rewritten in terms of the annihilation and creation operators in the form
\begin{equation}
\omega\left(\hat{a}^\dagger\hat{a}+\frac{1}{2}\right)\psi(\alpha) = \frac{1}{2}\left(\frac{\varepsilon}{J}+4\right)\psi(\alpha)\,.
\end{equation}
The lattice periodicity is taken into account by imposing that the Bloch wavefunction transform in the proper way under translations, namely one has
\begin{equation}
\psi_{n\vc{k}}(\vc{r}_{\vc{i}\alpha}) = \frac{1}{\sqrt{N_c}}\sum_s e^{i(k_xi_x+Qk_ys)a}\varphi_n\left(\alpha + Qi_y-Qs-\frac{Qk_xa}{2\pi}\right)\,.
\end{equation}
The above wavefunctions satisfy the properties required by Bloch theorem, namely $\psi_{n\vc{k}}(\vc{r}_{\vc{i}\alpha}+a\hat{\vc{x}}) = e^{ik_xa}\psi_{n\vc{k}}(\vc{r}_{\vc{i}\alpha})$ and $\psi_{n\vc{k}}(\vc{r}_{\vc{i}\alpha}+a Q\hat{\vc{y}}) = e^{iQk_ya}\psi_{n\vc{k}}(\vc{r}_{\vc{i}\alpha})$. In order to ease the notation in the following we label the positions of the lattice sites in the square lattice in the more natural way $\vc{r}_{\vc{i}} = a(i_x\hat{\vc{x}}+i_y\hat{\vc{y}})$, rather than $\vc{r}_{\vc{i}\alpha} = a(i_x\hat{\vc{x}}+(\alpha+Qi_y)\hat{\vc{y}})$.
Checking that the above wavefunctions are orthogonal and normalized is a good way to get a better feeling of the approximations that have been made (the number of lattice sites is $N_{\rm s} = L^2$ with $L = QM$ and the therefore then number of unit cells is $N_{\rm c} = N_{\rm s}/Q=QM^2$)
\begin{equation}
\begin{split}
\sum_{\vc{j}}\psi^*_{n\vc{k}'}(\vc{r}_{\vc{j}})\psi_{n\vc{k}}(\vc{r}_{\vc{j}}) &= \frac{1}{QM^2}\sum_{s,s'}\sum_{j_x,j_y}
e^{i(k_x-k_x')j_xa}e^{iQ(k_ys-k_y's')a}\times \\
&\times \varphi_n\left(j_y-Qs-\frac{Qk_xa}{2\pi}\right)\varphi_n\left(j_y-Qs'-\frac{Qk_x'a}{2\pi}\right) \\ &= \frac{\delta_{k_x,k_x'}}{M}\sum_{s,s'}\sum_{j_y}
e^{iQ(k_ys-k_y's')a}\varphi_n\left(j_y-Qs-\frac{Qk_xa}{2\pi}\right)\varphi_n\left(j_y-Qs'-\frac{Qk_xa}{2\pi}\right) \\
&= \frac{\delta_{k_x,k_x'}}{M}\sum_{s}\sum_{j_y}
e^{iQ(k_y-k_y')sa}\varphi^2_n\left(j_y-Qs-\frac{Qk_xa}{2\pi}\right) = \frac{\delta_{k_x,k_x'}}{M}\sum_{s}e^{iQ(k_y-k_y')sa} \\ &= \delta_{k_x,k_x'}\delta_{k_y,k_y'}\,.
\end{split}
\end{equation}
We have neglected terms where $s\neq s'$ since the harmonic oscillator wavefunctions are well localized inside the large ($Q$ sites) magnetic unit cell and their overlap with wavefunctions in neighbouring magnetic unit cell is exponentially small. Moreover we approximate the sum over $j_y$ with the integral $\sum_{j_y}\varphi_n^2(j_y)\approx \int dx\, \varphi_n^2(x)=1$, which leads to an error of the same order as it can be checked with the Euler-MacLaurin formula. The harmonic oscillator wavefunctions are normalized in the usual way. The \textit{periodic} Bloch functions $g_{n\vc{k}}(j_y)$ are thus (the same as Eq.~(25) in the main text with a different labelling)
\begin{equation}\label{eq:periodic_Bloch}
g_{n\vc{k}}(j_y) = \sum_s e^{-ik_y(j_y-Qs)a} \varphi_n\left(j_y-Qs-\frac{Qk_xa}{2\pi}\right)\,.
\end{equation}
We can now calculate $\mathcal{B}_{ij}(\vc{k})$ within this approximation for the periodic Bloch functions. We consider first the quantity $\bar{\mathcal{G}}_{\vc{k}}^\dagger (\partial_{k_j}\bar{\mathcal{G}}_{\vc{k}})$ that enters the second term in parenthesis in Eq.~(\ref{eq:M-def})
\begin{equation}\label{eq:overlap1}
\begin{split}
\bar{\mathcal{G}}_{\vc{k}}^\dagger (\partial_{k_x}\bar{\mathcal{G}}_{\vc{k}}) &= \sum_{j_y=1}^Q g_{\bar{n}\vc{k}}^*(j_y)\partial_{k_x}g_{\bar{n}\vc{k}}(j_y) \\ &= \sum_{j_y=1}^Q\sum_{s,s'}e^{-iQ(s-s')k_y}\varphi_{\bar{n}}\left(j_y-Qs-\frac{Qk_xa}{2\pi}\right)\partial_{k_x}\varphi_{\bar{n}}\left(j_y-Qs'-\frac{Qk_xa}{2\pi}\right) \\ 
&= \sum_{j_y=1}^Q\sum_s\varphi_{\bar{n}}\left(j_y-Qs-\frac{Qk_xa}{2\pi}\right)\partial_{k_x}\varphi_{\bar{n}}\left(j_y-Qs-\frac{Qk_xa}{2\pi}\right) \\
&=  -\frac{Qa}{2\pi}\sum_{j_y}\varphi_{\bar{n}}\left(j_y\right)\sqrt{\frac{\omega}{2}}(\hat{a}-\hat{a}^\dagger)\varphi_{\bar{n}}(j_y) \\ &= -\frac{Qa}{2\pi}\sqrt{\frac{\omega}{2}}\sum_{j_y}\varphi_{\bar{n}}\left(j_y\right)\left(\sqrt{\bar{n}}\varphi_{\bar{n}-1}(j_y)-\sqrt{\bar{n}+1}\varphi_{\bar{n}+1}(j_y)\right)=0\,.
\end{split}
\end{equation}
We have combined the sum over $j_y = 1, \dots,Q$ and the sum over $s$ in a single sum over all integer $j_y$. This sum over $j_y$ has then been approximated by an integral and the overlap of orthogonal functions $\varphi_n$ carried out as usual leading to a zero result. One can check that the choice of the periodic Bloch functions in Eq.~(\ref{eq:periodic_Bloch}) leads to $\bar{\mathcal{G}}_{\vc{k}}^\dagger (\partial_{k_y}\bar{\mathcal{G}}_{\vc{k}})\neq 0$. However we can perform a gauge transformation of the form $g_{n\vc{k}} \to e^{iQk_xk_ya^2/(2\pi)}g_{n\vc{k}}$. Going through the calculation along the same line of Eq.~(\ref{eq:overlap1}) one obtains something proportional to $\sum_{j_y}j_y\varphi_{\bar{n}}^2(j_y) \approx \int dx\,x\varphi_{\bar{n}}^2(x) = 0$.
We will make use of this gauge choice when convenient.

First we calculate $\mathcal{B}_{xy}(\vc{k})$. We only need to evaluate $\mathrm{Tr}\left[(\partial_{k_x}\bar{\mathcal{G}}_{\vc{k}}^\dagger)(\partial_{k_y}\bar{\mathcal{G}}_{\vc{k}})\right]$ since the second term in parenthesis in Eq.~(\ref{eq:M-def}) is zero according to Eq.~(\ref{eq:overlap1})
\begin{equation}
\begin{split}
&\frac{1}{2}\mathcal{B}_{xy}(\vc{k}) = \mathrm{Tr}\left[(\partial_{k_x}\bar{\mathcal{G}}_{\vc{k}}^\dagger)(\partial_{k_y}\bar{\mathcal{G}}_{\vc{k}})\right] = \sum_{j_y=1}^Q \partial_{k_x}g_{\bar{n}\vc{k}}^*(j_y)\partial_{k_y}g_{\bar{n}\vc{k}}(j_y) \\ 
&= \sum_{j_y=1}^Q \sum_{s,s'} \partial_{k_x}\left[e^{ik_y(j_y-Qs)a}
\varphi_{\bar{n}}\left(j_y-Qs-\frac{Qk_xa}{2\pi}\right)\right]\partial_{k_y}
\left[e^{-ik_y(j_y-Qs')a}
\varphi_{\bar{n}}\left(j_y-Qs'-\frac{Qk_xa}{2\pi}\right)\right] \\
&=\sum_{j_y} \partial_{k_x}\left[e^{ik_yj_ya}
\varphi_{\bar{n}}\left(j_y-\frac{Qk_xa}{2\pi}\right)\right]\partial_{k_y}
\left[e^{-ik_yj_ya}\varphi_{\bar{n}}\left(j_y-\frac{Qk_xa}{2\pi}\right)\right] \\
&= \frac{iQa^2}{2\pi}\sum_{j_y}\varphi_{\bar{n}}'\left(j_y-\frac{Qk_xa}{2\pi}\right) j_y \varphi_{\bar{n}}\left(j_y-\frac{Qk_xa}{2\pi}\right) \\
&= \frac{iQa^2}{2\pi}\sum_{j_y}\left[\sqrt{\frac{\omega}{2}}\left(\hat{a}-\hat{a}^\dagger\right)
\varphi_{\bar{n}}\left(j_y-\frac{Qk_xa}{2\pi}\right)\right]\left[\left(\frac{1}{\sqrt{2\omega}}\left(\hat{a}+\hat{a}^\dagger\right)+\frac{Qk_xa}{2\pi} \right)\varphi_{\bar{n}}\left(j_y-\frac{Qk_xa}{2\pi}\right)\right]\\
&= \frac{iQa^2}{4\pi}\sum_{j_y}\left[\sqrt{\bar{n}}\varphi_{\bar{n}-1}(j_y)-
\sqrt{\bar{n}+1}\varphi_{\bar{n}+1}(j_y)\right]\left[\sqrt{\bar{n}}\varphi_{\bar{n}-1}(j_y)+
\sqrt{\bar{n}+1}\varphi_{\bar{n}+1}(j_y)\right] = -\frac{iQa^2}{4\pi}\,.
\end{split}
\end{equation}
Next we evaluate $\mathcal{B}_{xx}(\vc{k})$. Again we need only need to evaluate $\mathrm{Tr}\left[(\partial_{k_x}\bar{\mathcal{G}}_{\vc{k}}^\dagger)(\partial_{k_x}\bar{\mathcal{G}}_{\vc{k}})\right]$ according to Eq.~(\ref{eq:overlap1})
\begin{equation}
\begin{split}
& \frac{1}{2}\mathcal{B}_{xx}(\vc{k}) = \mathrm{Tr}\left[(\partial_{k_x}\bar{\mathcal{G}}_{\vc{k}}^\dagger)(\partial_{k_x}\bar{\mathcal{G}}_{\vc{k}})\right] = \sum_{j_y=1}^Q \partial_{k_x}g_{\bar{n}\vc{k}}^*(j_y)\partial_{k_x}g_{\bar{n}\vc{k}}(j_y) \\
&=\sum_{j_y} \partial_{k_x}
\varphi_{\bar{n}}\left(j_y-\frac{Qk_xa}{2\pi}\right)\partial_{k_x}
\varphi_{\bar{n}}\left(j_y-\frac{Qk_xa}{2\pi}\right) \\
&= \left(\frac{Qa}{2\pi}\right)^2\sum_{j_y}\left[\sqrt{\frac{\omega}{2}}\left(\hat{a}-\hat{a}^\dagger\right)\varphi_{\bar{n}}\left(j_y\right)\right] \left[\sqrt{\frac{\omega}{2}}\left(\hat{a}-\hat{a}^\dagger\right) \varphi_{\bar{n}}\left(j_y\right)\right] \\
&= \frac{Qa^2}{4\pi}\sum_{j_y}\left[\sqrt{\bar{n}}\varphi_{\bar{n}-1}(j_y)-
\sqrt{\bar{n}+1}\varphi_{\bar{n}+1}(j_y)\right]\left[\sqrt{\bar{n}}\varphi_{\bar{n}-1}(j_y)-
\sqrt{\bar{n}+1}\varphi_{\bar{n}+1}(j_y)\right] \\ 
&= \frac{Qa^2}{4\pi}(2\bar{n}+1)\,.
\end{split}
\end{equation}
For $\mathcal{B}_{yy}(\vc{k})$ we choose the modified gauge that ensures the vanishing of the second term in parenthesis in Eq.~(\ref{eq:M-def})
\begin{equation}
\begin{split}
&\frac{1}{2}\mathcal{B}_{yy}(\vc{k}) = \mathrm{Tr}\left[(\partial_{k_y}\bar{\mathcal{G}}_{\vc{k}}^\dagger)(\partial_{k_y}\bar{\mathcal{G}}_{\vc{k}})\right] = \sum_{j_y=1}^Q \partial_{k_y}g_{\bar{n}\vc{k}}^*(j_y)\partial_{k_y}g_{\bar{n}\vc{k}}(j_y) \\
&=\sum_{j_y} \partial_{k_y}\left[e^{ik_y\left(j_y-\frac{Qk_xa}{2\pi}\right)a}
\varphi_{\bar{n}}\left(j_y-\frac{Qk_xa}{2\pi}\right)\right]\partial_{k_y}
\left[e^{-ik_y\left(j_y-\frac{Qk_xa}{2\pi}\right)a}\varphi_{\bar{n}}\left(j_y-\frac{Qk_xa}{2\pi}\right)\right] \\
&=a^2\sum_{j_y} \left[\left(j_y-\frac{Qk_xa}{2\pi}\right)\varphi_{\bar{n}}\left(j_y-\frac{Qk_xa}{2\pi}\right)\right]
\left[\left(j_y-\frac{Qk_xa}{2\pi}\right)\varphi_{\bar{n}}\left(j_y-\frac{Qk_xa}{2\pi}\right)\right] \\
&= a^2\sum_{j_y}\left[\frac{1}{\sqrt{2\omega}}\left(\hat{a}+\hat{a}^\dagger\right)\varphi_{\bar{n}}\left(j_y\right)\right] \left[\frac{1}{\sqrt{2\omega}}\left(\hat{a}+\hat{a}^\dagger\right) \varphi_{\bar{n}}\left(j_y\right)\right] \\
&= \frac{Qa^2}{4\pi}\sum_{j_y}\left[\sqrt{\bar{n}}\varphi_{\bar{n}-1}(j_y)+
\sqrt{\bar{n}+1}\varphi_{\bar{n}+1}(j_y)\right]\left[\sqrt{\bar{n}}\varphi_{\bar{n}-1}(j_y)+
\sqrt{\bar{n}+1}\varphi_{\bar{n}+1}(j_y)\right] \\
&= \frac{Qa^2}{4\pi}(2\bar{n}+1)\,.
\end{split}
\end{equation}
The quantum geometric tensor $\mathcal{B}_{ij}$ is constant over the Brillouin zone and the calculation of $\mathcal{M}$ is trivial (the integration is over the full Brillouin zone $k_x = [-\pi/a,\pi/a],\,k_x = [-\pi/(Qa),\pi/(Qa)]$)
\begin{equation}\label{eq:M-Landau}
\begin{split}
\mathcal{M} = \frac{1}{2\pi} \int d^2\vc{k}\,\mathcal{B}(\vc{k}) = \frac{Qa^2}{(2\pi)^2}\int_{\rm B.Z.} d\vc{k}\,
\begin{pmatrix}
2\bar{n}+1 & -i \\
i & 2\bar{n}+ 1\
\end{pmatrix}= \begin{pmatrix}
2\bar{n}+1 & -i \\
i & 2\bar{n}+ 1
\end{pmatrix}\,.
\end{split}
\end{equation}
We obtain the result that for a Landau level the Chern number is $|C|=1$ and that the superfluid density in the $\bar{n}$-th Landau level is proportional to $2\bar{n}+1$. In particular for the lowest Landau level the bound in Eq.~(23) in the main text is saturated.

\subsection*{Supplementary Note 5: Bound on the localization functional for Wannier functions}
\label{sec:bound_loc_func}

Marzari and Vanderbilt consider the following functional that can be minimized numerically in the matrix $U_{\vc{k}}$ in Eq.~(\ref{eq:definition}) in order to construct maximally localized  Wannier functions\ucit{Marzari:1997}{6} (we consider the 2D case for simplicity)
\begin{gather}
F = \sum_{\alpha}\left[\bra{\vc{0}\alpha}r^2\ket{\vc{0}\alpha} -\left|\bra{\vc{0}\alpha}\vc{r}\ket{\vc{0}\alpha}\right|^2\right]\,,\\
\bra{\vc{0}\alpha}r^2\ket{\vc{0}\alpha} = \int_\Omega d^2\vc{r}\,|w_\alpha(\vc{r})|^2r^2
  = \frac{A_\Omega}{(2\pi)^2}\int_{B.Z.} d^2\vc{k}\, \braket{\bm{\nabla}_{\vc{k}}g_{\alpha\vc{k}}}
  {\bm{\nabla}_{\vc{k}}g_{\alpha\vc{k}}} \,,\label{eq:ident1}\\
\bra{\vc{i}\alpha}\vc{r}\ket{\vc{0}\beta} = \int_\Omega d^2\vc{r}\,w^*_\alpha(\vc{r}-\vc{r}_{\vc{i}})w_\beta(\vc{r})\vc{r}
= i\frac{A_\Omega}{(2\pi)^2}\int_{B.Z.} d^2\vc{k}\,e^{i\vc{k}\cdot\vc{r}_{\vc{i}}}\braket{g_{\alpha\vc{k}}}{\bm{\nabla}_{\vc{k}}g_{\beta\vc{k}}} \,.\label{eq:ident2}
\end{gather}
$A_\Omega$ is the area of the unit cell, and the periodic Bloch functions indexed by $\alpha$ instead of $n$ are obtained by the transformation $g_{\alpha\vc{k}} = \sum_{n}[U_{\vc{k}}]_{\alpha,n}g_{n\vc{k}}$. We have also defined the ket $\ket{\bm{\nabla}_{\vc{k}}g_{\alpha\vc{k}}}$ as $\braket{\vc{r}}{\bm{\nabla}_{\vc{k}}g_{\alpha\vc{k}}} = \bm{\nabla}_{\vc{k}}g_{\alpha\vc{k}}(\vc{r})$ in the identities~(\ref{eq:ident1})-(\ref{eq:ident2}) which are derived in Ref.~\ncit{Marzari:1997}{6}.
The above functional can obviously be interpreted as the sum of the spreads of the Wannier functions. The localization functional can be split into two contributions
\begin{gather}
F = F_{I} + \widetilde{F}\,,\\
F_{I} = \sum_\alpha\left[ \bra{\vc{0}\alpha}r^2\ket{\vc{0}\alpha} -\sum_{\vc{i}\beta}\left|\bra{\vc{i}\beta}\vc{r}\ket{\vc{0}\alpha}\right|^2 \right]\,,\\
\widetilde{F} = \sum_{\substack{\vc{i}\beta\\ \vc{i}\beta \neq \vc{0}\alpha}}\left|\bra{\vc{i}\beta}\vc{r}\ket{\vc{0}\alpha}\right|^2\,.
\end{gather}
The first contribution $F_{I}$ is gauge invariant in the sense that it is independent of the unitary matrix $U_{\vc{k}}$ and it is interesting to provide an expression for $F_{I}$ in terms of the Bloch functions $g_{\alpha\vc{k}}$ using the identities~(\ref{eq:ident1})-(\ref{eq:ident2})
\begin{equation}
\begin{split}
F_{I} &= \sum_i\sum_\alpha\bigg[\frac{A_\Omega}{(2\pi)^2}\int_{B.Z.} d^2\vc{k}\, \braket{\partial_{k_i}g_{\alpha\vc{k}}}
  {\partial_{k_i}g_{\alpha\vc{k}}}\\ &\qquad-\sum_{\vc{i}\beta}\left(\frac{A_\Omega}{(2\pi)^2}\right)^2\int_{B.Z.} d^2\vc{k}_1\int_{B.Z.} d^2\vc{k}_2\,e^{i(\vc{k}_1-\vc{k}_2)\cdot\vc{r}_{\vc{i}}}\braket{\partial_{k_i}g_{\alpha\vc{k}_2}}{g_{\beta\vc{k}_2}}\braket{g_{\beta\vc{k}_1}}{\partial_{k_i}g_{\alpha\vc{k}_1}}\bigg]\,\\
  &=  \frac{A_\Omega}{(2\pi)^2}\sum_i\int_{B.Z.} d^2\vc{k}\sum_\alpha\bigg[\braket{\partial_{k_i}g_{\alpha\vc{k}}}
  {\partial_{k_i}g_{\alpha\vc{k}}} + \sum_\beta\braket{g_{\alpha\vc{k}}}{\partial_{k_i}g_{\beta\vc{k}}}
  \braket{g_{\beta\vc{k}}}{\partial_{k_i}g_{\alpha\vc{k}}}\bigg]\\
 &= \frac{A_\Omega}{(2\pi)^2}\sum_i\int_{B.Z.} d^2\vc{k} \,\big[
 \mathrm{Tr}[(\partial_i\mathcal{\bar G}^\dagger_{\vc{k}})(\partial_i\mathcal{\bar G}_{\vc{k}})]+ \mathrm{Tr}[\mathcal{\bar G}_{\vc{k}}^\dagger
 \left(\partial_i\mathcal{\bar G}_{\vc{k}}\right)
 \mathcal{\bar G}_{\vc{k}}^\dagger\left(\partial_i\mathcal{\bar G}_{\vc{k}}\right)]\big] \\
 &= \frac{A_\Omega}{(2\pi)^2}\sum_i\int_{B.Z.} d^2\vc{k}\, \frac{1}{2}\mathcal{B}_{ii}(\vc{k}) = \frac{A_\Omega}{2\pi} \,\frac{1}{2}\mathrm{Tr}[\mathcal{M}]\,.
\end{split}
\end{equation}
In the last line we have used the definition of quantum geometric tensor $\mathcal{B}_{ij}(\vc{k})$ and of the invariant matrix $\mathcal{M}$ [see Eq.~(\ref{eq:M-def})]. In particular the matrix $[\mathcal{\bar G}_{\vc{k}}]_{\vc{r},\alpha} = g_{\alpha\vc{k}}(\vc{r})$ is analogous to the matrix $\mathcal{\bar G}_{\vc{k}}$ as defined in Eq.~(\ref{eq:Gbar-def}) and used in the main text with the only difference that the index $\vc{r}$ is continuous and has to been integrated on the whole unit cell when the matrix product is performed. Using the arithmetic mean-geometric mean inequality $(\lambda_1+\lambda_2)/2  \geq \sqrt{\lambda_1\lambda_2}$ for the eigenvalues of the positive matrix $\mathcal{M}$ and indicating with $\mathcal{M}^{\rm R}$ the real part of $\mathcal{M}$ and with $\mathcal{M}^{\rm I}$ the imaginary part, one has the following chain of equalities/inequalities
\begin{equation}
\frac{1}{2}\mathrm{Tr}[\mathcal{M}] = \frac{1}{2}\mathrm{Tr}[\mathcal{M}^{\rm R}] \geq 
\sqrt{\mathrm{det}[\mathcal{M}^{\rm R}]} \geq \sqrt{\mathrm{det}[\mathcal{M}^{\rm I}]} = |C|\,. 
\end{equation}
The second inequality is Eq.~(23) in the main text. Thus the Chern number of the composite band provides a bound from below of the localization functional, namely the result stated in the main text
\begin{equation}
F \geq \frac{A_\Omega}{2\pi}|C|\,.
\end{equation}

\subsection*{Supplementary Note 6: Berezinsky-Kosterlitz-Thouless transition}
\label{sec:bkt}

The low energy fluctuations of the order parameter phase $\theta(\vc{r})$ have an energy cost given by the Hamiltonian of the classical 2D XY model
\begin{equation}
\mathcal{H}_{\rm XY} = \frac{\Upsilon}{2}\int d^2\vc{r}\,\big(\bm{\nabla}\theta(\vc{r})\big)^2\,.
\end{equation}
The \qql stiffness\qqr  $\Upsilon$ is in fact related to the superfluid weight by the relation $\Upsilon = D_{\rm s}\hbar^2/4$ where the factor 4 that appears in this last equation is due to the fact that in our notation $\bm{\nabla}\theta(\vc{r}) = 2\vc{q}$ for a constant phase gradient. At the Berezinsky-Kosterlitz-Thouless transition temperature $T_{\rm BKT}$ the vortex unbinding due to thermal fluctuations leads to the destruction of the phase order and thus of superfluidity. The BKT critical temperature can be obtained by the universal relation\ucit{Nelson:1977}{7} $\Upsilon(T_{\rm BKT}) /(k_{\rm B}T_{\rm BKT})= 2/\pi$. We would like to estimate the BKT critical temperature with the finite temperature result obtained for the superfluid weight~(\ref{eq:superfluid_weight_flat}). For simplicity only the half filling case $\nu = 1/2$ is considered. The zero temperature value of the energy gap $\Delta(T=0) = Un_\phi/2 = 2k_{\rm B}T_{\rm c}$ is taken as the energy scale. The temperature-dependence of the energy gap is then obtained by the equation (compare with Eq.~(\ref{eq:Ebar}) and~(\ref{eq:delta_eq}))
\begin{equation}\label{eq:DeltaT}
\frac{\Delta(T)}{\Delta(0)} = \tanh\left(\frac{T_{\rm c}}{T}\frac{\Delta(T)}{\Delta(0)}\right)\,,
\end{equation}
while the stiffness is, from Eq.~(\ref{eq:superfluid_weight_flat}),
\begin{equation}
\frac{\Upsilon_{\bar{n}}(T)}{\Delta(0)} = \frac{2\bar{n}+1}{4\pi}\frac{\Delta^2(T)}{\Delta^2(0)} \,.
\end{equation}
The dependence of the stiffness on the Landau level number $\bar{n}$ has been introduced. Then the equation that needs to be solved in order to find the BKT critical temperature is 
\begin{equation}\label{eq:TBKT}
\frac{2\bar{n}+1}{4}\frac{\Delta^2(T_{\rm BKT})}{\Delta^2(0)} = \frac{T_{\rm BKT}}{T_{\rm c}}\,.
\end{equation}
Eq.~(\ref{eq:DeltaT}) and~(\ref{eq:TBKT}) can be solved numerically and the result is shown in Supplementary Figure~\ref{fig:BKT}. The BKT temperatures for $\bar{n} = 0,1,2$ are approximately $T_{\rm BKT} \approx 0.25,\;0.61,\; 0.75\,T_{\rm c}$, respectively. This justifies our assertion in the main text that the mean-field temperature is in fact a good estimate of the real temperature of the superconducting transition in the lowest bands of the Harper model.

\subsection*{Supplementary References}
\begin{itemize}
\item[1.] \label{Schnyder:2008}
Schnyder, A. P.,  Ryu, S., Furusaki, A. \& Ludwig, A. W. W., Classification of topological insulators and superconductors in three spatial dimensions, \href{http://journals.aps.org/prb/abstract/10.1103/PhysRevB.78.195125}{Phys. Rev. B \textbf{78}, 195125 (2008)}.
\item[2.]\label{Taylor:2006}
Taylor, E., Griffin, A., Fukushima, N. \& Ohashi, Y., Pairing fluctuations and the superfluid density through the BCS-BEC crossover, \href{http://journals.aps.org/pra/abstract/10.1103/PhysRevA.74.063626}{Phys. Rev. A \textbf{74}, 063626 (2006)}.
\item[3.]\label{Grosso_Book}
Grosso G., \& Pastori Parravicini, G., \textit{Solid State Physics}, 2nd ed., Elsevier (2014).
\item[4.]\label{Brouder:2007}
Brouder, C., Panati, G., Calandra, M., Mourougane, C. \& Marzari, N., Exponential Localization of Wannier Functions in Insulators, \href{http://dx.doi.org/10.1103/PhysRevLett.98.046402}
{Phys. Rev. Lett. \textbf{98}, 046402 (2007)}.
\item[5.]\label{Harper:2014}
Harper, F., Simon, S. H. \& Roy, R., Perturbative approach to flat Chern bands in the Hofstadter model, \href{http://journals.aps.org/prb/abstract/10.1103/PhysRevB.90.075104}{Phys. Rev. B \textbf{90}, 075104 (2014)}.
\item[6.]\label{Marzari:1997}
Marzari, N.,  \& Vanderbilt, D., Maximally localized generalized Wannier functions for composite energy bands, \href{http://journals.aps.org/prb/abstract/10.1103/PhysRevB.56.12847}
{Phys. Rev. B \textbf{56}, 12847-12865 (1997)}.
\item[7.]\label{Nelson:1977}
Nelson, D. R. \& Kosterlitz, J. M., Universal Jump in the Superfluid Density of Two-Dimensional Superfluids, \href{http://journals.aps.org/prl/abstract/10.1103/PhysRevLett.39.1201}
{Phys. Rev. Lett. \textbf{39}, 1201-1204 (1977)}.
\end{itemize}